\PassOptionsToPackage{table}{xcolor}
\documentclass[acmsmall,screen]{acmart}

\AtBeginDocument{%
  }

\setcopyright{acmlicensed}
\copyrightyear{2026}

\usepackage{array}
\usepackage{booktabs}
\usepackage{caption}
\usepackage{float}
\usepackage{geometry}
\usepackage{graphicx}
\usepackage{longtable}
\usepackage{multirow}
\usepackage{siunitx}
\usepackage{subfig}
\usepackage{tikz}
\usetikzlibrary{positioning,fit}
\usepackage{url}
\usepackage{xspace}
\usepackage[normalem]{ulem}
\usepackage{amsmath}

\newcolumntype{L}[1]{>{\raggedright\arraybackslash}p{#1}}
\newcolumntype{P}[1]{>{\centering\arraybackslash}p{#1}}

\author{Arpita Wadhwa}
\orcid{0009-0009-4260-6645}
\affiliation{%
  \institution{Harvard University}
  \city{Cambridge}
  \country{USA}}
\email{arpitawadhwa@hks.harvard.edu}

\author{Aditya Vashistha}
\orcid{0000-0001-5693-3326}
\affiliation{%
  \institution{Cornell University}
  \city{Ithaca}
  \country{USA}}
\email{adityav@cornell.edu}

\author{Mohit Jain}
\orcid{0000-0002-7106-164X}
\affiliation{%
  \institution{Microsoft Research}
  \city{Bangalore}
  \country{India}}
\email{mohja@microsoft.com}

 \newcommand{\chiadd}[1]{\textcolor{black}{#1}}
 \newcommand{\chidelete}[1]{}

\newcommand{\todo}[1]{}
\newcommand{\mobilehciadd}[1]{\textcolor{black}{#1}}
\newcommand{\mobilehcidelete}[1]{}
\newcommand{\chiaddnew}[1]{\textcolor{black}{#1}}
\newcommand{\chideletenew}[1]{}
\newcommand{\jankaari}{\textit{Info-Bot}\xspace}
\newcommand{\asha}{\textit{ASHA-Bot}\xspace}
\newcommand{\ashasaheli}{\textit{ASHA-Friend-Bot}\xspace}
\newcommand{\swasthyavibhag}{\textit{Health-Dept-Bot}\xspace}

\begin{document}

\title{Trade-offs in Social-Norm Framings for Health Chatbots: Balancing Trust and Preference}

\begin{abstract}

AI-driven chatbots are increasingly being used to support community health workers (CHWs) in developing regions. Yet little is known about how cultural frameworks in chatbot design shape trust in collectivist contexts where decisions are rarely made in isolation. This paper examines how CHWs in rural India responded to chatbot\chiadd{-interface}s that delivered identical health content but varied in one specific cultural lever: \emph{social norms}.
Through a mixed-methods study with 61 ASHAs who compared four normative framings: neutral, descriptive, narrative identity, and injunctive authority, we (1) analyze how framings influence preferences and trust and (2) compare effects in low- and high-ambiguity scenarios. The results show that narrative framings were most preferred but encouraged overreliance, while authority framings were least preferred yet supported calibrated trust.
We conclude with design recommendations for dynamic framing strategies that adapt to context and argue for \textit{calibrated trust}
as a critical evaluation metric for safe, culturally-grounded AI.
\end{abstract}


\setcopyright{cc}
\setcctype{by}
\acmJournal{PACMHCI}
\acmYear{2026} \acmVolume{10} \acmNumber{5} \acmArticle{MHCI8049}
\acmMonth{8} \acmDOI{10.1145/3821679}


\maketitle

\section{Introduction}

Community Health Workers (CHWs) are the backbone of primary care in many low- and middle-income countries, providing last-mile services in rural and underserved areas~\cite{baqui2008}. In rural India, nearly one million CHWs, called Accredited Social Health Activists (ASHAs), support maternal and neonatal health. ASHAs are typically women with high-school education, recruited from their own communities to provide essential care, often with limited training, minimal supervision, and delayed access to expert advice~\cite{Shresth2024,Scott2019,Dhaliwal2025}. When guidance is unclear or delayed, the consequences can be severe: a missed referral for preeclampsia, a delayed newborn checkup, or an unnecessary hospitalization can threaten lives and erode trust in public health programs. 

Conversational AI, delivered through chatbots or messaging apps, has emerged as a promising tool to address these gaps by providing on-demand answers to time-sensitive health questions, supplementing knowledge, and bridging language barriers \cite{ashabot-chi25, Car2020, Okolo2021, Okolo2024}. But in high-stakes domains like healthcare, the legitimacy of AI depends not only on \textit{what} information it provides but also on \textit{how} that information is framed in ways that resonate with ASHAs' cultural context and institutional realities. A chatbot that ``sounds like a peer'' may invite comfort and openness, whereas one that ``sounds like an authority'' may convey credibility. For ASHAs, such framings are not cosmetic---they can determine whether AI advice is trusted, contested, or dismissed, directly influencing health decisions and patient outcomes. 

\chidelete{This challenge is particularly acute in collectivist contexts such as India, where health decisions are rarely made in isolation but are negotiated within families, peer networks, and community norms. ASHAs }\chiadd{ASHAs} also constantly balance government-approved health protocols with \chidelete{these }social realities, translating institutional rules into guidance that households will accept.
Their work is fundamentally persuasive: they cannot command compliance, but must navigate culture, norms, and power relations to encourage care-seeking and behavior change.
In this context, chatbot advice must not only be \textit{accurate}\chiadd{,} but\chidelete{ must} also feel credible and culturally aligned for them to follow it~\cite{kok2016}. 
Normative cues are central to this credibility.
Theories of social norms show that guidance is more likely to be followed when it reflect\chiadd{s} what peers do (descriptive norms) or what authority figures endorse (injunctive norms)~\cite{Cialdini1991,Bicchieri2006}. Designing culturally aligned AI therefore requires attention not only to the content of advice but also to the normative framings through which that advice is delivered.

\mobilehciadd{In practice, however, AI tools for CHWs have been largely designed around technical priorities, including retrieval accuracy, topic coverage, and reducing hallucinations~\cite{Car2020}, assuming users evaluate outputs on informational merits alone. ASHAs, however, are frontline workers, not model engineers. They encounter chatbots mid-fieldwork on smartphones, and assess trustworthiness through social and cultural cues (who is speaking, with what authority, in what voice) more readily than through technical signals~\cite{Verdezoto2021}. The result is a gap between how CHW-facing chatbots are built (information-first) and how CHWs decide whether to act on their advice (framing-first). As LLM-powered tools scale across the Global South~\cite{ashabot-chi25}, this gap carries direct safety implications: choices builders treat as cosmetic, such as bot name, persona, or source of authority, may shape whether advice is followed, questioned, or misapplied.}


Yet, despite growing calls for culturally aligned AI \cite{cacm2023culturalAI,varuvel_dennison_designing_2026}, we lack empirical evidence on how these normative framings in chatbot design shape ASHAs' trust and preferences in practice. 
Do peer-based framings feel more relatable and preferred by ASHAs, but risk \chideletenew{blind acceptance}\chiaddnew{overreliance}? Do authority-based framings feel less preferred, but promote greater trust through institutional legitimacy? Answers to these questions determine how chatbots will be used, trusted, and acted upon in high-stakes health contexts where AI overreliance is a known risk~\cite{Okolo2021,Okolo2024}.
Therefore, we ask: 
\begin{itemize}
    \item[\textbf{RQ1:}] How do different normative framings of a health chatbot influence ASHAs' preferences and trust in its advice?
    \item[\textbf{RQ2:}]How do the effects of these framings on trust differ between low-ambiguity and high-ambiguity scenarios?  
\end{itemize}

To answer \chidelete{our }\chiadd{these} questions, we conducted a mixed-method evaluation of four WhatsApp-based chatbot\chiadd{-interface} designs for ASHAs, each delivering identical health content but with a distinct normative framing: descriptive norm (\chiadd{what} other ASHAs often do), narrative identity norm (peer stories), injunctive authority norm (voice of the health department), and a neutral baseline.
We recruited 61 ASHAs in rural Rajasthan, India.
The study consisted of structured preference and trust tasks with in-depth interviews. \chiadd{We first gather ASHAs' comparative preferences across 4 chatbot-interfaces. Following which we assessed trust in}\chidelete{In the trust task, we examined} both low-ambiguity scenarios (with clear protocol-based answers) and high-ambiguity scenarios (where guidance exists but often clashes with on-the-ground realities).

Our findings show that normative framings strongly shaped both preferences and trust.
The narrative identity chatbot\chiadd{-interface} that spoke through stories of other ASHAs maximized comfort and relatability, and hence was the most preferred. 
Yet it also encouraged \textit{overreliance}, with ASHAs often trusting even its 
incorrect advice. By contrast, the injunctive authority framing---where the chatbot\chiadd{-interface} spoke in the formal voice of the health department---was the least preferred but reduced overreliance. Its bureaucratic tone prompted workers to pause, cross-check, and resist advice that felt inconsistent, producing more \textit{calibrated trust}.
Under increased ambiguity, framing effects intensified: narrative framings amplified \chideletenew{blind acceptance}\chiaddnew{overreliance}, while authority framings sharpened discernment between correct and incorrect advice.

Based on our findings, we argue for moving away from the blind maximization of “cultural alignment” in AI. Instead, we propose treating social norms as a tunable design lever aimed at calibrated trust. Specifically, we recommend \textit{dynamic norm framings} that switch between norms depending on the scenario, and \textit{feasibility-aware advice} by chatbot\chiadd{-interface}s in scenarios where established protocols can conflict with local constraints. We also call for evaluation of AI-powered chatbot\chiadd{-interface}s in terms of calibrated trust, rather than merely user preference or \chideletenew{blind trust}\chiaddnew{overreliance}, so that systems can be designed and optimized accordingly. In summary, this paper makes the following contributions to HCI:

\begin{itemize}
    \item[\textbf{1.}] It provides one of the first empirical studies of how social norms framings---descriptive, narrative, injunctive authority, and neutral---shape ASHAs' preferences and trust when interacting with chatbot-interfaces. In doing so, it \chideletenew{also }extends norm-based health communication research\chiaddnew{, which has largely examined broadcast-style SMS interventions with null or modest effects~\cite{Papakonstantinou2025},} \chideletenew{beyond broadcast-style SMS interventions}to a dialogic, conversational modality \chiaddnew{where normative cues may operate differently}.
    \item[\textbf{2.}] It advances design implications for culturally aligned AI in collectivist health systems, highlighting how narrative identity framings can foster resonance but risk overreliance, while injunctive authority framings can dampen engagement yet safeguard discernment.
\end{itemize}
Together, these contributions extend debates on cultural alignment in AI from theory to evidence, offering both methodological tools and design guidance for systems that are simultaneously resonant, reliable, and accountable.

\section{Related Work}
We situate our work within three strands of literature. First, cultural critiques of technology and AI establish that while culture clearly matters for the design and adoption of AI systems, there has been limited investigation into the extent of this influence and the potential risks of overemphasizing cultural cues. Second, we review research on social norms in health behavior and HCI to show how norms have been leveraged as powerful levers for behavior change, but note a gap in empirical work that systematically compares multiple norm framings in AI-mediated contexts. Third, we turn to studies on AI for CHWs to highlight how questions of trust and overreliance shape adoption in practice. 

\subsection{Cultural Critiques of AI and Technology}
Scholarship in HCI and ICTD has long challenged universalist paradigms that treat technologies as culture-neutral. Early ICTD debates emphasized that technological interventions fail when divorced from local contexts~\cite{Toyama2015}, while postcolonial computing highlighted how design practices reinscribe Western epistemologies and marginalize alternative ways of knowing~\cite{Irani2010,Philip2012}. \citet{Dourish2012} further note that ubiquitous computing discourses often erase the material realities of the Global South, privileging abstract Western ideals over lived practices. These critiques underscore that technologies are never neutral but always situated, raising concerns about systems transplanted wholesale across cultural borders. 

Contemporary AI research echoes these concerns. Large language models, presented as universal, encode narrow linguistic and cultural assumptions~\cite{Bender2021}. \citet{Birhane2021} critiques the field’s reliance on agnostic individual rationality, which renders local and contextual knowledge invisible. \citet{Mohamed2020} proposed Decolonial AI, advocating accountability rooted in non-Western epistemologies, while \citet{Sambasivan2021} show that technological interventions can fail when cultural contexts are ignored. In response, HCI and ICTD researchers have developed culturally grounded approaches, designing interfaces that leverage local idioms, metaphors, and social logics to foster usability and legitimacy~\cite{Medhi2012, farmchat_imwut2018,varuvel_dennison_designing_2026}. In AI ethics~\cite{Greene2019,Madaio2020}, scholars argue that systems must be evaluated not only on accuracy but also on cultural resonance, especially in domains such as health where legitimacy is critical. Yet, while much of the literature insists that culture matters, less attention has been paid to how to design culturally aligned AI systems ~\cite{varuvel_dennison_designing_2026} and whether maximizing cultural cues can itself carry unintended risks. 
Our study extends this conversation by empirically examining how social norms---a key driver of culture---shape frontline workers’ preference and trust in AI systems.

\subsection{Social Norms in Health Behavior and HCI}
Social norms are shared expectations of what is typical or appropriate. \chiaddnew{Literature}\chideletenew{Social and cultural theory} commonly distinguishes three types: \emph{descriptive} norms (what most people do), \emph{narrative identity} norms (stories\chideletenew{ and exemplars} that define what "people like me" do), and \emph{injunctive} norms (what authorities approve or disapprove)~\cite{Cialdini1991,Bicchieri2006,Hofstede2001,GreenBrock2000}. Norm-based interventions have been shown to influence health decisions across diverse contexts, including vaccination uptake~\cite{Tankard2016}, safer sexual practices~\cite{Perkins2003,Latkin2005}, sanitation~\cite{Goldstein2008}, and harassment reduction\chideletenew{ in schools}~\cite{Paluck2010}.\chideletenew{ with appeals to norms proving most effective when they resonate with local cultural logics~\cite{Chung2023}.} Building on \chiaddnew{this}\chideletenew{these insights}, HCI researchers have embedded normative cues in digital systems\chideletenew{. Personal informatics tools that display peer comparisons have encouraged} to encourage healthier behaviors~\cite{Munson2012}\chideletenew{and in ICTD contexts, norm-based designs have} and promote sanitation, safe water use, and maternal health practices in resource-constrained settings~\cite{Medhi2012,Derenzi2017}. \chideletenew{Studies of digital agents further show that normative cues, such as peer identity and authority, shape perceptions of trust and legitimacy~\cite{Seeger2021,Zhou2019,Araujo2018} }However, three important limitations constrain this literature.

First, evidence on norm-based messaging is decidedly mixed. \chiaddnew{Experiments}\chideletenew{Prior experiments, including} in India~\cite{Rimal2024, Sedlander2021} and Colombia~\cite{Maldonado}\chiaddnew{,} find that descriptive norms often yield null or counterproductive effects in low-adoption contexts, inadvertently reinforcing the undesirable status quo, while injunctive norms produce only modest improvements~\cite{Cialdini2003,Schultz2007}. A\chideletenew{recent} meta-analysis of 89 randomized\chideletenew{controlled} trials concludes that norm messages have largely null average effects across SMS, posters, and print media~\cite{Papakonstantinou2025}. \chiaddnew{One explanation is that such broadcast formats lack interpersonal cues needed for norms to feel personally relevant.} \chiaddnew{Conversational AI may change this.} Unlike SMS interventions, which deliver unidirectional broadcast messages, chatbots can embody social identities through names, avatars, and linguistic style. They engage users in turn-taking dialogue\chiaddnew{, creating an illusion of interpersonal exchange and tailoring responses to individual queries. These affordances could amplify normative effects by making social cues feel personally directed---a peer speaking to the user, not a system messaging everyone~\cite{Araujo2018,Feine2019}.} 
\mobilehciadd{This perspective echoes prior MobileHCI work, which shows that mobile messaging platforms such as WhatsApp are not neutral delivery channels but socially embedded infrastructures carrying expectations of responsiveness, credibility, and peer presence~\cite{church2013, loerakker2025}. Work on frontline mobile health systems further shows that technologies are evaluated in situ, under real-world constraints of mobility, workflow, and social context~\cite{dell2014}.}
\mobilehciadd{Building on this, HCI work has begun to examine how interface-level cues shape user trust in health AI~\cite{schmidmaier2025, Zhan2024}. For instance, Kollerup et al.~\cite{Kollerup2024} show that textual cues manipulating an AI dermatologist's perceived ability and integrity shift users' self-reported trust even when the underlying information is held constant. 
Mendel et al.~\cite{Mendel2024} show that the advice source (doctor vs. AI) mediates users' willingness to disclose health information to remote monitoring systems. Together, these studies establish that source framing in health AI is not cosmetic but consequentially shapes user behavior. What remains under-examined is how these effects play out when the source cues are drawn from distinct normative traditions (peer, narrative, or institutional), and whether they differentially shape trust.}

\chiaddnew{Second, prior work targets patients or the general public~\cite{Cialdini1991, Bicchieri2006, Chung2023, Papakonstantinou2025, Sedlander2021, Rimal2024}, but less is known about how normative framings operate when recipients are health workers. This matters because workers face fundamentally different decision contexts. While parents deciding whether to vaccinate their child might respond to "the Health Department recommends it,"  an ASHA deciding whether to persist when a family resists must balance official protocol against social feasibility, family relationships, supervisor expectations, and observed practices of other ASHAs. She operates within a professional hierarchy, faces accountability for outcomes, and must translate institutional rules into guidance households will accept. Norms that persuade patients may not help workers navigate these competing demands. Our study addresses this gap by examining norm-based design in a worker-facing context, where stakes include both individual decisions and downstream effects on patient care.}

Third, most studies test only a single norm framing against a control\chiaddnew{,}\chideletenew{ condition,} leaving open questions about how different framings compare\chideletenew{ to one another}. Even fewer examine the risk of overreliance, where users accept AI guidance too readily, even when\chideletenew{it is} incorrect. Our outcome measure diverges from prior work\chiaddnew{:}\chideletenew{ in this respect:} most norm-based interventions evaluate real-world health behaviors such as vaccination uptake or clinic attendance\chiaddnew{.}\chideletenew{. In contrast, we} \chiaddnew{We} evaluate decision behavior within the human-AI exchange~\cite{Maldonado,Rimal2024,Hallsworth2016}\chiaddnew{,}\chideletenew{. Specifically, we measure} \chiaddnew{measuring} \emph{calibrated trust}: whether users appropriately follow correct advice while resisting incorrect advice---a safety-oriented metric aligned with\chideletenew{ongoing} concerns about appropriate reliance in human-AI collaboration~\cite{Bansal2021,Bucinca2021}.
\chiaddnew{By simultaneously addressing these three gaps, our work sits at the intersection of health communication, human-AI interaction, and ICTD as one of the first empirical tests of normative framing effects in conversational AI for frontline health workers.}\chideletenew{Our study addresses these three gaps simultaneously by testing the effect of multiple normative cues in worker-facing contexts. This positions our work at the intersection of health communication, human-AI interaction, and ICTD, offering one of the firsts empirical test of normative framing effects in conversational AI for frontline health workers.}

\subsection{AI for ASHAs and Community Health Workers}
CHWs play a critical role in delivering primary health services in low- and middle-income countries. In India, ASHAs connect households with the public health system, supporting maternal and child health, vaccination mobilization, and community health education. \mobilehciadd{Ethnographic accounts have documented the substantial relational and articulation work that ASHAs perform alongside these formal duties, navigating community trust, supervisor expectations, and household dynamics as part of everyday care delivery~\cite{Verdezoto2021}.} ICTD research has thus explored digital tools to assist CHWs in domains such as data collection~\cite{Medhi2012, medhi_nordichi12}, peer learning~\cite{mobilizing_chi08}, knowledge dissemination~\cite{vashistha_examining_2017,vashistha_mobile_2016}, and performance feedback~\cite{Kaphle2015,derenzi_closing_2016}. These studies demonstrate that CHWs welcome tools that enhance their credibility and efficiency, but remain cautious about technologies that increase monitoring or administrative burden~\cite{Dell2012}.

AI-enabled systems are beginning to transform this landscape. Recent LLM-powered solutions, such as ASHABot~\cite{ashabot-chi25}, support ASHAs in addressing their information needs. Studies show that diagnostic apps have been seen as empowering because they provide rapid access to medical knowledge, but also burdensome due to concerns about errors and accountability \cite{Okolo2021}. 
Later work found that CHWs valued AI outputs because they legitimized their advice to patients, even when the explanations were opaque~\cite{Okolo2024,solano-kamaiko_explorable_2024}. 
While AI can bolster confidence, it also introduces risks of overreliance, where workers defer to algorithmic outputs and discount their own expertise~\cite{Goddard2012, Cabitza2017, Lyell2017}. 
AI decision support for screening, referral, and medication management has also been piloted \cite{Praveen2014, Labrique2018} with CHWs and primary-care clinicians in decision-support systems for specific conditions. 
Systematic reviews, however, caution that many such systems neglect the cultural and social dynamics that shape CHWs’ decisions, limiting their adoption and trustworthiness~\cite{Agarwal2015}. Qualitative studies further emphasize that for AI to be useful, it must be comprehensible, verifiable, and culturally legitimate, aligning with both government protocols and community expectations~\cite{Okolo2021}.
\mobilehciadd{Taken together, this body of work has examined CHW-facing AI primarily through the lens of accuracy, explainability, and information retrieval, treating the chatbot as a knowledge resource whose value is determined largely by what it knows. Less attention has been paid to how the same information is received depending on who the chatbot appears to be speaking as: a peer, an institution, a friend, or a neutral system. This is an important area to study because, as the qualitative literature above suggests, CHWs often decide whether to follow advice based on the social legitimacy of its source as much as its technical provenance.}

Our study builds on \mobilehciadd{and extends} this literature by shifting focus from accuracy and usability to cultural framing. We examine whether the normative framing of AI advice—as peer-like, narrative, or authoritative—alters how ASHAs prefer and trust its guidance. By embedding these framings into otherwise identical WhatsApp chatbots, we isolate how cultural cues shape both preferences and calibrated reliance. In doing so, we extend debates about AI for CHWs beyond performance metrics, offering one of the first empirical examinations of how cultural alignment can both enable and complicate responsible adoption in global health contexts.

\section{Methodology}
\label{methods}
To answer our research questions, we conducted a mixed-methods study with ASHAs in rural Rajasthan, India. \chiadd{The study was conducted in close collaboration with \textit{Khushi Baby}, an on-ground organization with more than a decade of experience training frontline workers on digital tools.} \chiadd{In this section,} we first describe how we operationalized normative framing into distinct chatbot\chidelete{s}\chiadd{-interface} designs, \mobilehciadd{validated them}, followed by our study protocol, recruitment and participant demographics, and ethical safeguards. Finally, we describe the collected data that result from structured quantitative tasks combined with qualitative interviews.

\begin{table*}
\centering
\caption{Theoretical differences in the social norms.}
\label{tab:norms-diff}
\Description{This table summarizes the theoretical differences among three types of social norms: descriptive, narrative identity, and injunctive. These norms vary in their definitions, underlying mechanisms, typical formats, and psychological triggers, as well as in the cultural contexts where they are most effective. Descriptive norms emphasize what most people are already doing through neutral, statistical framing; narrative identity norms leverage peer stories to foster empathy and identification; and injunctive norms draw authority from formal rules and institutions, promoting compliance through directive, rule-based communication. Further details are provided in the Methodology section.}
\footnotesize
\setlength{\tabcolsep}{4pt}
\rowcolors{2}{gray!20}{white}
\begin{tabular}{@{}L{0.12\textwidth} L{0.27\textwidth} L{0.27\textwidth} L{0.27\textwidth}@{}}
\hline
\textbf{} &
\textbf{Descriptive Norms~\cite{Cialdini1991, Goldstein2008, Markus1991}} &
\textbf{Narrative Identity Norms~\cite{McAdamsMcLean2013, GreenBrock2000, HinyardKreuter2007}} &
\textbf{Injunctive Norms~\cite{Cialdini1991, Hofstede2001}}
\\ \hline
\textbf{Definition} &
  Highlights what most peers are already doing &
  Embeds guidance in first-person stories from relatable peers &
  Communicates what trusted authorities (e.g., govt, supervisors) approve or expect \\
\textbf{Mechanism} &
  \textit{Social proof}: behavior appears safe and appropriate because it is common &
  \textit{Identification \& modeling}: users see themselves in the peer's story and actions &
  \textit{Legitimacy \& accountability}: credibility derives from institutional authority \\
\textbf{Typical Format} &
  Peer statistics or generalized statements &
  Mini-testimonials or narrative vignettes &
  Rule-based, guideline-oriented phrasing \\
\textbf{Psychological Trigger} &
  Conformity, uncertainty reduction, perceived safety in alignment &
  Empathy, resonance, and social modeling &
  Clarity, procedural confidence, compliance with formal rules \\
\textbf{Cultural Role} &
  Effective in collectivist settings where group practice legitimizes action &
  Salient in relational cultures (e.g., South Asia) where shared identity builds trust &
  Dominant in hierarchical contexts where legitimacy flows from authority \\
\textbf{Tone} &
  Neutral, observational, factual &
  Warm, peer-like, conversational &
  Formal, directive, institutional \\ \hline
\end{tabular}
\end{table*}
\begin{table*}
\centering
\caption{Chatbot framings and sample responses to the question: ``\textit{When should the BCG vaccination be given?}''}
\Description{This table illustrates how the same factual guidance can be framed differently across chatbot designs. The Info-Bot provides a neutral, factual response, while the ASHA-Bot frames the advice through descriptive norms, highlighting what most ASHAs do. The ASHA-Friend-Bot uses a narrative identity norm, embedding the advice in a relatable peer story. Finally, the Health-Dept-Bot invokes injunctive norms by referencing official guidelines and institutional authority.}
\label{tab:sample-bots}
\footnotesize
\setlength{\tabcolsep}{5pt}
\begin{tabular}{@{}P{0.235\textwidth} P{0.235\textwidth} P{0.235\textwidth} P{0.235\textwidth}@{}}
\hline
\textbf{\jankaari} (Baseline) &
  \textbf{\asha} &
  \textbf{\ashasaheli} &
  \textbf{\swasthyavibhag} \\ \hline
\includegraphics[width=0.10\textwidth]{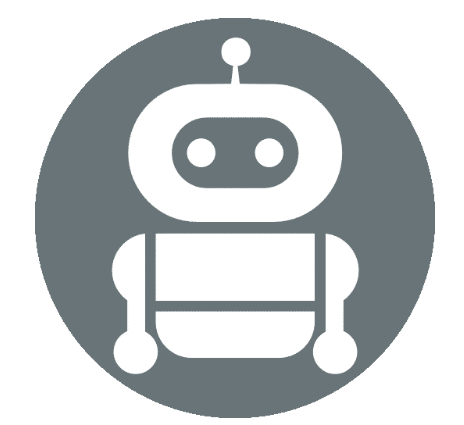} &
  \includegraphics[width=0.10\textwidth]{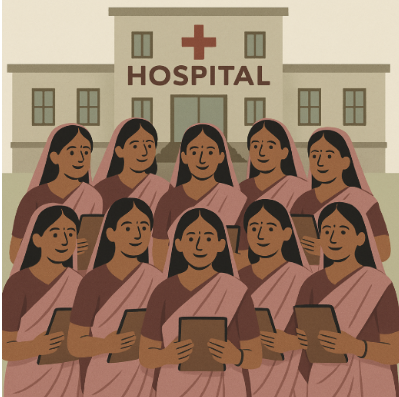} &
  \includegraphics[width=0.10\textwidth]{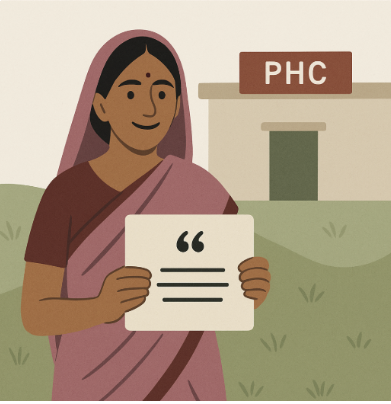} &
  \includegraphics[width=0.10\textwidth]{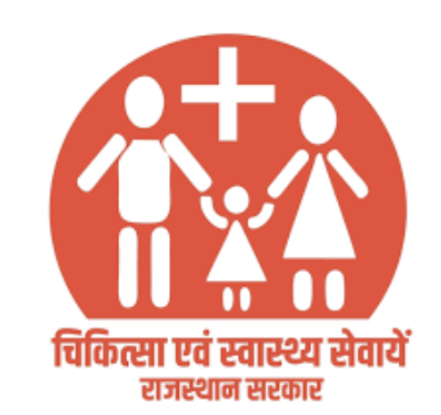} \\
\addlinespace
The BCG vaccine should be given at birth. If it is missed at birth, it can still be given up to one year of age. &
  84\% ASHAs in your block give BCG at birth; if missed, still within one year. &
  Radhika, an ASHA in your area, said BCG should be given at birth, or within the first year if missed. &
  As per Health Department guidelines, BCG must be given at birth. If missed, it should be administered as soon as possible, up to one year of age. \\ \hline
\end{tabular}
\end{table*}

\subsection{Operationalizing Social Norms}
Social psychology and cultural theory \cite{Cialdini1991,Bicchieri2006} identify three primary types of norms that guide human behavior: descriptive norm, narrative identity, and injunctive authority. We translated these 
\chiadd{framings} into three corresponding \chidelete{chatbot framings }\chiadd{chatbot-interfaces}, each signaling its normative role through the bot's name, display picture, and message style (including onboarding prompts, nudges, and responses to ASHA queries). To isolate the effect of framing, we also designed a neutral baseline \chidelete{bot }\chiadd{chatbot-interface}.

Table \ref{tab:norms-diff} summarizes these framings, and Table~\ref{tab:sample-bots} illustrates their contrasting answers to the same BCG vaccination query.
Importantly, all four chatbot\chiadd{-interfaces} were identical in informational content and \chidelete{interface }\chiadd{presentation}; only the normative framing varied.

\textbf{Neutral Baseline: \jankaari.} Information-Bot delivered informational content without any social cues. Its display picture was a plain robot (Table~\ref{tab:sample-bots}), and its tone was impersonal. This baseline enabled us to assess the additive effect of normative framings.

\textbf{Descriptive Norms: \asha.}
Descriptive norms emphasize what most peers typically do. The mechanism is social proof, with the cognitive pull of conformity: “if others like me do this, it must be correct.” \asha, with a display picture of a group of ASHAs, framed responses in aggregated terms (e.g., “\textit{84\% ASHAs in your block ...}”), reinforcing that following peer behavior is the correct choice \chiadd{~\cite{Cialdini1991, Goldstein2008, Markus1991}}.

\textbf{Narrative Identity Norms: \ashasaheli.}  
Narrative identity norms embed guidance within stories from relatable peers. The mechanism is identification and social modeling, with the cognitive pull of resonance: “she is like me, I know her, so I can follow what she did.” \ashasaheli
conveyed companionship, using an ASHA holding a testimony card as the display picture.
Its responses were story-based and
voiced through named peers (e.g., “\textit{Radhika, an ASHA in your area ...}”) \chiadd{~\cite{McAdamsMcLean2013, GreenBrock2000, HinyardKreuter2007}}.

\textbf{Injunctive Authority Norms: \swasthyavibhag.}  
Injunctive norms communicate what authorities prescribe. The mechanism is legitimacy and accountability, with the cognitive pull of compliance: “if the Health Department directs this, I am expected to do it.” \swasthyavibhag (“Health-Department-Bot”) displayed the health department emblem as its display picture and used formal, directive phrasing (e.g., “\textit{As per Health Department guidelines ...}”) \chiadd{~\cite{Cialdini1991, Hofstede2001}}.

\chiadd{For all chatbot-interface variations, we created single-turn Q\&A responses (static mocks, shown in Figure~\ref{fig:sample_p1}) rather than deploying a live system. \mobilehcidelete{This prevented two common issues in live LLM-based deployments: models generating slightly different answers to identical queries and unpredictable response delays that can affect user judgments. It also enabled a structured comparative setup
where participants evaluated multiple controlled versions of the same query, which was necessary to estimate marginal effects of each normative framing under equivalent conditions.}This approach follows established HCI practice of using non-interactive prototypes (mockups) to assess interaction hypotheses prior to field trials~\cite{Cox_2022, Yun_2025, Metzger_2024}.} \mobilehciadd{Thus, our setup is closer to an ablation than a naturalistic field deployment. We keep the advice and structure the same, and only change the type of framing (descriptive, narrative-identity, or injunctive). Stripping away conversational dynamics lets us attribute differences in ASHA responses to the framing manipulation itself. Using live LLM-based systems would make this harder to do. Models can give slightly different answers to the same question, so participants might end up comparing different responses rather than different framings. Response times can also vary, and delays alone can affect how trustworthy or reliable a system feels. In addition, differences in how conversational the bots are (e.g., how much they elaborate or clarify) could introduce other unintended cues. By controlling these factors through static mocks, we keep the comparison focused on framing.}

\mobilehciadd{\subsection{Field-Based Validation of Chatbot-Interfaces}}

\mobilehciadd{Next, we validated our translation from theoretical norm constructs to interface designs through a three-stage process grounded in ASHAs' interpretations. First, one co-author drafted candidate text for each bot's name, onboarding message, daily nudge, and Q\&A pairs, based on formal definitions of descriptive, narrative-identity, and injunctive norms~\cite{Cialdini1991,Bicchieri2006,GreenBrock2000,Hofstede2001}. These drafts were then refined through team discussion. Second, four NGO field staff reviewed all materials to assess local legibility (e.g., whether an ASHA in rural Rajasthan would interpret ``Health Department guidelines'' as an institutional voice) 
and to ensure that the three bots remained distinct while presenting identical informational content. This process led to several revisions. For example, the descriptive bot's phrasing was localized from ``\textit{many ASHAs do X}'' to ``\textit{ASHAs in your block do X},'' because field staff noted that block-level framing carries more social weight. Third, we conducted a pilot study  with five ASHAs who were not included in the final sample. We asked each participant to describe who they believed was speaking through each bot. All participants identified Health-Dept-Bot as ``the government,'' ASHA-Friend-Bot as ``a fellow ASHA sharing her own story,'' and ASHA-Bot as ``what other ASHAs around here are doing,'' confirming that the intended framings were consistently understood.} \mobilehciadd{Taken together, these steps helped validate that the chatbot-interfaces were distinct, and were being identified with the intended framings.}



\subsection{Study Design}
\label{subsec:study-design}

\begin{figure*}[]
    \centering
    \includegraphics[width=0.75\linewidth]{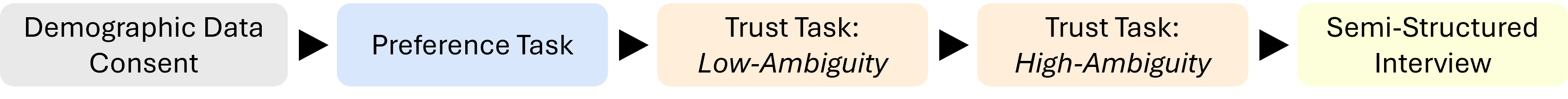}
    \caption{Study procedure.}
    \Description{The study proceeded in four stages. First, participants provided demographic data and consent. They then completed the Preference Task, where they interacted with chatbot\chiadd{-interface} variants. Next, they undertook two Trust Tasks: one with low-ambiguity cases and another with high-ambiguity cases. Finally, each session concluded with a semi-structured interview to capture qualitative insights.}
    \label{fig:study-flow}
\end{figure*}

The study comprised three structured components---(1) preference tasks, (2) low-ambiguity trust tasks, and (3) high-ambiguity trust tasks---followed by a semi-structured interview, as shown in Figure~\ref{fig:study-flow}. This design allowed us to examine not only which chatbot\chiadd{-interface} framings ASHAs preferred, but also whether these framings supported \emph{calibrated trust}, i.e., following correct advice while resisting incorrect recommendations. 

\chiadd{We employed a mixed-factorial design following established HCI methodology~\cite{li2023, rae2024}, with the preference tasks served as a within-subject factor and the trust tasks as a ~\mobilehciadd{partial} between-subject factor. More details are provided below. Task types, question placement, and chatbot pair assignments were fully counterbalanced across participants; individual questions were randomly assigned without replacement. To ensure accuracy and contextual validity, four NGO field staff reviewed all prompts, questions, and chatbot responses, ensuring their alignment with common ASHA information needs, local health contexts, and official training materials.} 

\begin{figure*}[]
    \centering
    \includegraphics[width=0.6\linewidth]{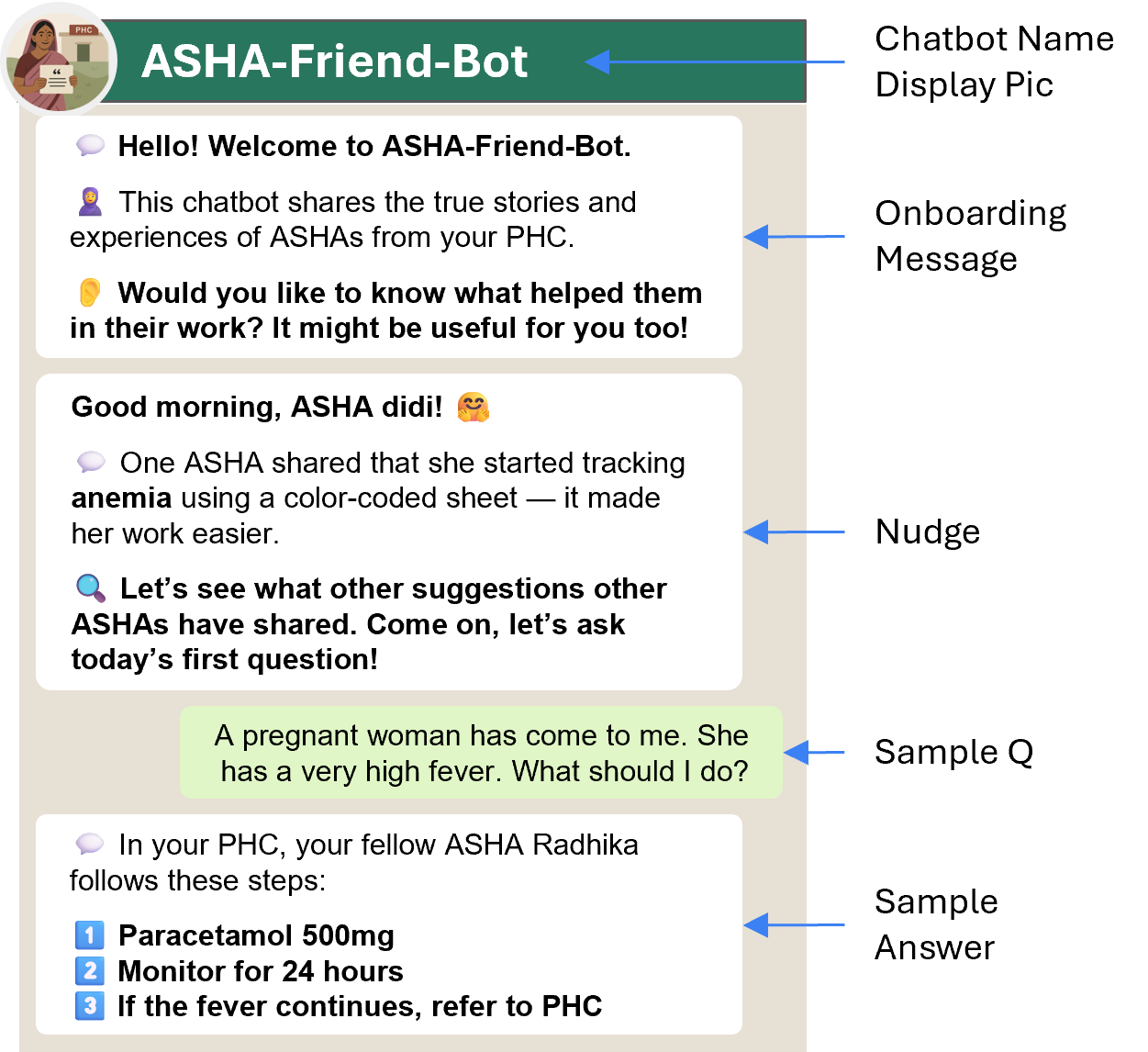}
    \caption{Illustration of the Preference Task, showing the interface for \ashasaheli.}
    \Description{Illustration of the interface for ASHA-Friend-Bot, used in the Preference Task. The interface displays four sequential chat bubbles: (1) an onboarding message introducing the bot and its purpose of sharing ASHA experiences; (2) a nudge framed as a peer story highlighting anemia tracking; (3) a sample user query about managing a pregnant woman with high fever; and (4) a sample bot response referencing peer practice, outlining three recommended steps (paracetamol, monitoring, referral).}
    \label{fig:sample_p1}
\end{figure*}

\textbf{Preference Tasks.}
\label{pref-tasks}
Each ASHA completed two preference comparisons. In each, she was shown two chatbot\chiadd{-interface}s side by side (randomly drawn from the four). The bot interface image displayed five consistent elements: (1) display picture, (2) chatbot name, (3) onboarding message, (4) a nudge, and (5) one sample Q\&A pair. ASHAs were asked to select the \chidelete{bot }\chiadd{interface} they preferred and explain why. Figure~\ref{fig:sample_p1} shows the interface for \ashasaheli.
The second task repeated the procedure with the remaining two chatbot\chiadd{-interface}s, ensuring that every ASHA compared all four framings while content was held constant. To minimize bias, initial \chidelete{chatbot }pairs were randomized across the six possible combinations, and their left/right placement was counterbalanced. 

\textbf{Trust Tasks.}
\label{trust-tasks}
Preference alone does not capture the trust ASHAs placed in a chatbot's advice. Trust is critical in healthcare contexts, as ASHAs face high-stakes situations (e.g., pregnancy complications, child illness) and the accuracy of guidance can directly affect health outcomes. For digital tools to be useful, ASHAs must trust them enough to act on their advice, yet misplaced or uncritical trust is risky, as overreliance may lead to harmful decisions. Thus, effective use requires discernment: knowing when to follow and when to override chatbot responses.

To evaluate trust, participants completed four trust tasks: two in low-ambiguity scenarios and two in high-ambiguity scenarios. In each, they reviewed \chiadd{static mock-ups of} a chatbot conversation (ASHA query + bot response) and indicated whether they would follow the advice (\emph{Yes/No}), explaining their reasoning. \chiadd{To measure trust calibration, we balanced the correctness of the chatbot outputs: half of the chatbot responses were correct and half incorrect.} \mobilehciadd{We used a partial-exposure design: each ASHA was randomly assigned to evaluate two of the four chatbot-interfaces in the low-ambiguity phase and the remaining two in high-ambiguity. This balances two constraints. A fully within-subjects design (16 scenarios per ASHA) would have led to fatigue, while a fully between-subjects design with 61 ASHAs across four conditions would be underpowered. The partial-exposure design preserves random assignment to bot $\times$ ambiguity conditions while keeping per-participant load manageable, and counterbalancing ensured each chatbot-interface received comparable exposure across the sample.}

\textit{Low-Ambiguity Trust Tasks.}  
These scenarios involved routine health questions with clear protocol-defined answers such as timing of BCG vaccination, exclusive breastfeeding (See Table~\ref{tab:amb-exemplar}).
For each participant, one bot from each earlier preference pair was randomly chosen, ensuring continuity and balanced exposure. To avoid question-specific effects, we created a set of five low-ambiguity questions, with both correct and incorrect responses.

\textit{High-Ambiguity Trust Tasks.}  
The same procedure was used for more complex scenarios where “correct” actions are less clear or contested such as family resistance to institutional delivery (See Table~\ref{tab:amb-exemplar}). \chiadd{Similar to the low-ambiguity tasks,} a separate set of five high-ambiguity questions was developed, using material from the official ASHA training guides, with both correct and incorrect answers. 

\mobilehciadd{All prompts remained grounded in real-world ASHA practice. Specifically, questions were drawn from the government-mandated ASHA training curriculum~\cite{ASHA_training_material}, high-ambiguity scenarios were identified by NGO field staff reflecting recurring tensions in actual practice (e.g., family resistance to institutional delivery), and four NGO staff reviewed all prompts for contextual realism. The presentation medium also matched ASHAs' existing habits, since WhatsApp is already their primary channel for communicating with peers and supervisors~\cite{Jamison2025}.} Additionally, across both ambiguity conditions, questions were randomly assigned without replacement, ensuring that no ASHA encountered the same question twice. Although individual questions were not perfectly balanced across chatbot\chiadd{-interface}s, their overall distribution was comparable. \chiadd{Additionally, each participant saw only two of the possible combinations (5 sample questions $\times$ 4 chatbot framings $\times$ 2 accuracy conditions), and was not informed that the underlying content was identical across chatbot-interfaces; hence, each scenario was treated as an independent evaluation. An entirely within-subjects design for the trust tasks would have required each ASHA to complete 16 scenarios, risking fatigue, carryover effects, and demand characteristics, as participants might infer the study's hypotheses from repeated exposure to all framings~\cite{Bucinca2021}. Our partial exposure approach, thus, balanced statistical power against these methodological threats. To check for potential carryover effects, we compared response patterns across task orders and found no systematic differences in preference or trust, suggesting that order effects were minimal.}

\textbf{Semi-Structured Interviews.}  
Each session concluded with a 25-minute interview probing participants’ reasoning---what shaped their preferences, how they decided to trust or override advice, the kinds of questions they would post to such systems, and how chatbot framings resonated with their everyday work. The interviews were conducted one-on-one by the first author in Hindi, audio-recorded with consent, and transcribed daily to minimize data loss.

\textbf{Procedure.}
Fieldwork took place over two weeks in August 2024. Each session was conducted one-on-one by the first author at the sub-center nearest each ASHA, to minimize travel burden. A pilot study revealed that group settings silenced quieter participants and encouraged conformity biases; hence, we adopted one-on-one study format.
Each session lasted \textasciitilde{}55 minutes: 10 minutes for introduction, consent, and demographic data collection; 20 minutes for structured tasks; and 25 minutes for the interview. The researcher explained the purpose and tasks of the study in simple terms, obtained written consent, and collected demographic information (age, education and work experience). The instructions were read aloud to accommodate literacy differences. All study materials were in Hindi and were administered by a researcher fluent in the language as well as familiar with the local sociocultural context. Participants were not compensated for this study, but received financial support from our partner NGO, \textit{Khushi Baby}, for a separate project deployment.

\begin{table*}
\centering
\caption{High-ambiguity task example. Q: \textit{A mother prefers institutional delivery but her family resists. How should the ASHA proceed?}}
\Description{This table shows how each chatbot framed responses to a high-ambiguity scenario: ``A mother prefers institutional delivery but her family resists. How should the ASHA proceed?'' Each chatbot provided both a correct and an incorrect response, aligned with its framing style. While the Info-Bot and Health-Dept-Bot emphasized guideline-based directives, the ASHA-Bot and ASHA-Friend-Bot embedded advice in peer practices and narratives. The task tested whether participants could discern the correct response.}
\label{tab:amb-exemplar}
\small
\setlength{\tabcolsep}{5pt}
\rowcolors{2}{gray!20}{white}
\begin{tabular}{@{}>{\raggedright\arraybackslash}p{0.12\textwidth} >{\raggedright\arraybackslash}p{0.40\textwidth} >{\raggedright\arraybackslash}p{0.40\textwidth}@{}}
\hline
\textbf{Chatbot} &
  \textbf{Correct response} &
  \textbf{Incorrect response} \\ \hline
\textit{Info-Bot} &
  Even if the family resists, continue counseling for institutional delivery. &
  If the family does not agree, stop persuading them. \\
\textit{ASHA-Bot} &
  In your block, 81\% of ASHAs continue to counsel for institutional delivery even when families hesitate. &
  In your block, 81\% of ASHAs accept the family's decision and stop counseling for institutional delivery. \\
\textit{ASHA-Friend-Bot} &
  An ASHA at your PHC shared that she firmly counseled families and supported institutional delivery. &
  An ASHA at your PHC shared that when the family resisted, she stopped counseling further. \\
\textit{Health-Dept-Bot} &
  According to Rajasthan Health Department guidelines, ASHAs should counsel firmly for institutional delivery. &
  According to Rajasthan Health Department guidelines, if the family refuses, ASHAs should not insist further. \\ \hline
\end{tabular}
\end{table*}

\subsection{Sample Size and Power Calculation}
\label{subsec:power}
We purposively recruited $N{=}61$ ASHAs. Eligibility required that ASHAs had not previously used any LLM-powered chatbots (e.g., ASHABot~\cite{ashabot-chi25}), to avoid contamination of treatment effects. Each ASHA completed two head-to-head \emph{preference} comparisons ($61\times 2 = 122$ tasks) and four \emph{trust} tasks (two low-ambiguity and two high-ambiguity, $61\times 4 = 244$ tasks). 
Because multiple tasks came from the same individual, the effective sample size was smaller than the raw task counts. Adjusting for clustered data, this yielded 100--110 effective observations for preference and 140--190 for trust.\chiaddnew{\footnote{See Appendix A for details on design-effect adjustment and power simulations.}}
\chideletenew{The effective observation counts were computed using the standard cluster design-effect adjustment to account for repeated measures from the same ASHA, calculated as:
$N_{\text{eff}} = \frac{N_{\text{raw}}}{1 + (n - 1)\rho}$,
where $n$ is the number of tasks completed per ASHA and $\rho$ is the intra-class correlation (ICC) across responses. We estimated plausible ICC values (0.10--0.20) using random-intercept models at the ASHA level. Power simulations~\cite{duflo2006toolkit} confirmed that this design achieved $\sim$80\% power to detect main framing effects of 12--15 percentage points at $\alpha{=}0.05$. These values are typical for clustered behavioral experiments and were sufficient to detect main framing effects (e.g., whether one chatbot framing was systematically preferred, whether ASHAs calibrated trust by following correct more often than incorrect advice), though not subtle subgroup interactions (e.g., framing $\times$ ambiguity $\times$ education).} To complement our quantitative findings, interviews provided qualitative depth, surfacing why workers responded differently to framings and how they understood trust in practice.

\subsection{Participants}
61 ASHAs (all females) with an average age of 40.7$\pm$6.5 years (range: 29--57) participated in the study. On average, they had 14.9$\pm$4.7 years of experience as ASHAs.
Educational attainment varied: 13.1\% had studied up to middle school or less, 68.8\% had completed middle- to high-school, and 18.1\% held an undergraduate degree or higher.

\subsection{Ethics and Positionality}
The study was approved by the Institutional Review Board at \textit{Microsoft Research India}. Written informed consent was obtained from all participants in Hindi, emphasizing voluntary participation and the right to withdraw without penalty. All data and transcripts were anonymized. Our positionality shaped both the design and interpretation of this study. 

The researchers, having prior fieldwork experience with ASHAs, collaborated closely with a local NGO partner to ensure the chatbot framings resonated with lived practice rather than abstract theory. At the same time, our institutional affiliations—as researchers from global academic and corporate research institutions—may have influenced how ASHAs perceived the authority of the study. We mitigated these dynamics by foregrounding the role of NGO fieldworkers, who were trusted intermediaries, and by framing the study as exploratory rather than evaluative of ASHA performance. We also acknowledge that, although the NGO partnership helped bridge cultural and institutional distance, as external researchers, our outsider perspective may still have constrained how we understood ASHAs’ narratives.

\chiadd{Another key ethical consideration in our design was how to introduce the necessary experimental manipulations in a responsible way. Studying the effects of different normative framings required varying who the chatbot\chiadd{-interface} ``spoke as'' and presenting both correct and incorrect responses, a well-established methodology in HCI for studying trust \cite{Cox_2022, Bucinca2021, Bansal2021}. To handle these design requirements ethically, we conducted a thorough debrief at the end of each session, informing participants that some chatbot-interface responses had been intentionally made inaccurate for research purposes. We also clarified that messages framed as coming from “other ASHAs” or “the Health Department” were simulated and not real. This ensured that participants did not leave with incorrect information or with mistaken beliefs about peer or institutional guidance.}


\subsection{Data Analysis}
\label{subsec:analysis}
Our analysis combined structured task outcomes with qualitative interviews.

First, from the 122 preference tasks, we examined the distribution of choices across framings. To test differences, we used chi-square tests and a Bradley--Terry model, which estimates odds ratios for each chatbot\chiadd{-interface} relative to the neutral baseline (\jankaari) and significance for head-to-head matchups. We also explored whether preferences varied by ASHA characteristics (age, education, experience) using logistic regressions with interaction terms.

Next, from the 244 trust tasks (balanced across low-/high-ambiguity and correct/incorrect answers), we modeled
whether ASHAs reported they would follow chatbot advice (Yes/No). To distinguish persuasiveness from discernment, we decomposed trust into four metrics:

\begin{itemize}
  \item \textsc{Overall Trust:} probability of following regardless of correctness.  
  $\text{Overall Trust} = \Pr(\text{Follow} \mid \text{All Answers})$.

  \item \textsc{Appropriate Trust:} probability of following when advice was correct.
  $\text{Appropriate Trust} = \Pr(\text{Follow} \mid \text{Correct Answers})$.

  \item \textsc{Overreliance\chiaddnew{\footnote{We use ``overreliance'' to denote the behavioral pattern of accepting incorrect AI advice---a safety-critical outcome in health contexts. This framing foregrounds calibration accuracy rather than the cognitive or social processes underlying trust decisions. We acknowledge that what manifests as overreliance may reflect situated trust transfer---the reasonable application of peer-trust heuristics to systems that mimic peer voice---rather than irrational deference. Section~\ref{sec:discussion} reflects on this interpretive tension and alternative approaches that can be employed in future work.}}:} probability of following when advice was incorrect.
  $\text{Overreliance} = \Pr(\text{Follow} \mid \text{Incorrect Answers})$.

  \item \textsc{Calibrated Trust:} difference between appropriate trust and overreliance. 
  $\text{Calibrated Trust} = \Pr(\text{Follow} \mid \\ \text{Correct Answers}) - \Pr(\text{Follow} \mid \text{Incorrect Answers})$.
\end{itemize}

We used logistic regressions to test (1) whether accuracy predicted trust (See Equation \ref{eq:model1}), (2) whether calibrated trust differed across chatbot\chiadd{-interfaces} (See Equation \ref{eq:model2}), and (3) whether demographics moderated these effects (See Equation \ref{eq:model3}). 

\mobilehciadd{Our design measures 
\emph{behavioral} trust, captured through reliance decisions in tasks where advice can be followed or overridden~\cite{Vereschak2021}. 
The dependent variable in every trust task is the binary action, ``Will you follow this chatbot's advice? (Yes/No),'' and all four metrics (Overall Trust, Appropriate Trust, Overreliance, Calibrated Trust) are computed from these decisions rather than self-reported ratings. We chose this approach for two reasons. First, in high-stakes health contexts, 
the critical outcome is whether advice is acted upon; behavioral measures capture trust as it manifests in action.
Second, behavioral measures are less vulnerable to social-desirability and acquiescence biases documented in prior HCI work in collectivist contexts, including among ASHAs~\cite{Dell2012}. Concretely, we operationalize a ``trust violation'' as following incorrect advice (overreliance) and ``calibrated trust'' as following correct advice while declining incorrect advice. Table~\ref{tab:amb-exemplar} shows correct and incorrect responses to the same high-ambiguity scenario across all four bots. Qualitative interviews complement these measures, but our core trust claims rest on the behavioral data.}

For qualitative findings, interview transcripts were analyzed using \mobilehciadd{Braun and Clarke's reflexive} thematic analysis~\cite{braun2006thematicanalysis}. 
\mobilehciadd{The first author conducted line-by-line inductive coding across all 61 transcripts, producing an initial codebook of 47 codes.
Through iterative discussions, all authors consolidated these into five higher-order themes: affective responses to framings, decision-making rationales for trust and override, spontaneous cross-framing comparisons, references to professional practice and supervisor relationships, and contextual constraints (e.g., distance, family dynamics, network availability).}  \mobilehciadd{Please refer to Appendix D for details on the qualitative coding process and our codebook.}

\section{Findings}
\label{sec:findings}

We begin by examining which normative framings ASHAs preferred and draw on qualitative insights to explain why some framings resonated more than others (Section~\ref{sec:preferences}). We then analyze how these framings shaped trust and overreliance, first in low-ambiguity tasks where correct answers were clear (Section~\ref{sec:low}) and then in high-ambiguity tasks where protocols conflicted with on-the-ground realities (Section~\ref{sec:high}). Throughout, we use the neutral \jankaari as the baseline for all comparisons.

\subsection{\chidelete{Chatbot}\chiadd{Chatbot-Interface} Preferences}
\label{sec:preferences}

\subsubsection{Overall Preference Distribution}

We begin by examining which chatbot\chiadd{-interface} ASHAs preferred in the pairwise choice task. A chi-square goodness-of-fit test confirmed that the observed distribution of choices deviated significantly from a uniform 25.0\% expectation ($\chi^2(3)=32.4$, $p<.001$), indicating that participants consistently favored certain chatbot\chiadd{-interface}s over others.

We then looked at the six possible head-to-head comparisons among the four chatbot\chiadd{-interface}s. The \ashasaheli was the overwhelming favorite, winning 53 of 61 comparisons (86.9\%). At the other extreme, the \swasthyavibhag was the least favored, winning only 11 comparisons (18.0\%). The \asha performed moderately well with 36 wins (59.0\%), while the neutral \jankaari won 22 comparisons (36.1\%). Table~\ref{tab:pairwise-results-tab} shows winner in each of these comparisons and the magnitude of each win.

In five of the six head-to-head comparisons, one \chidelete{bot}\chiadd{interface} was clearly preferred over the other, with the difference large enough to be statistically significant (Table~\ref{tab:pairwise-results-tab}). The only exception was the comparison between the \asha and the \swasthyavibhag. Here, preferences were more evenly split (59.0\% vs.\ 41.0\%), and the difference was not statistically significant. This suggests that while narrative framing generally outperformed the others, the authority framing still held some appeal when put directly against descriptive framing.

Although these pairwise tests tell us which \chidelete{designs}\chiadd{interfaces} were significantly preferred in direct comparisons, they did not tell us how strong those preferences were relative to a common benchmark. To address this, we estimated a Bradley-Terry model\mobilehciadd{~\cite{bradleyterry1952}}, using the neutral \jankaari as the baseline. \mobilehciadd{This model is a standard statistical approach widely used in HCI for analyzing pairwise comparisons~\cite{chiang2024, perez2019, Mantiuk2012}.
It estimates a latent ``strength'' for each option such that the probability of one option being chosen over another follows a logistic function of their difference. In our case,}
it puts all interfaces on the same scale (against the baseline), so we can compare not just which one was preferred more, but how strongly each framing was preferred. Our analysis found that \ashasaheli was 7.5$\times$ more likely to be chosen (Table~\ref{tab:bt-model}), \asha 2.1$\times$ more likely, while \swasthyavibhag was only 0.46$\times$ as likely as compared to the baseline.

These results confirm the clear and systematic ordering of chatbot\chiadd{-interface} preferences: 
\begin{center}
\textbf{\ashasaheli $>$ \asha $>$ \jankaari $>$ \swasthyavibhag}
\end{center}


\begin{figure*}[]
    \centering
    \begin{minipage}{0.4\textwidth}
        \centering
        \captionof{table}{Chatbot comparison using Bradley-Terry model in the Preference Task.}
        \Description{This table presents results from a Bradley-Terry model estimating ASHAs' stated chatbot preferences, with Info-Bot as the baseline. Peer framings significantly increased preference: ASHA-Bot doubled the odds of being chosen (OR = 2.1, p = .036), while ASHA-Friend-Bot had the strongest effect, making it over seven times more likely to be preferred (OR = 7.4, p = .001). By contrast, the authoritative Health-Dept-Bot framing was significantly less preferred than baseline (OR = 0.4, p = .043). These findings indicate that ASHAs gravitated most strongly toward peer-like framings, particularly those embedding personal narratives.}
        \label{tab:bt-model}
        \footnotesize
        \setlength{\tabcolsep}{4pt}
        \begin{tabular}{@{}lcccc@{}}
            \toprule
            \textbf{Bot (vs.\ Baseline)} & $\boldsymbol{\beta}$ & $\begin{array}{c}\text{Odds}\\\text{Ratio}\end{array}$ & \textbf{95\% CI} $(\beta)$ & $p$ \\
            \midrule
            \textbf{\jankaari} (baseline) & $0.0$ & $1.0\times$ & -- & -- \\
            \textbf{\asha} & $+0.7$ & $2.1\times$ & [0.05, 1.48] & .036 \\
            \textbf{\ashasaheli} & $+2.0$ & $7.4\times$ & [1.14, 2.88] & .001 \\
            \textbf{\swasthyavibhag} & $-0.7$ & $0.4\times$ & [$-1.54$, $-0.03$] & .043 \\
            \bottomrule
        \end{tabular}
    \end{minipage}
    \hfill
    \begin{minipage}{0.48\textwidth}
        \centering
        \captionof{table}{Pairwise head-to-head chi-square tests.}
        \Description{This table reports results from pairwise comparisons of ASHAs' stated chatbot preferences, with Holm correction for multiple comparisons. ASHA-Friend-Bot was the clear overall favorite, winning over all other framings with large margins (84-90\%, all p < .01). ASHA-Bot was also significantly preferred over Info-Bot (p < .0001), though not significantly different from Health-Dept-Bot. In contrast, Health-Dept-Bot was consistently the least preferred option, losing decisively to both Info-Bot and ASHA-Friend-Bot. These results show that ASHAs most strongly favored peer-like narrative framings, while authoritative framings were least appealing.}
        \label{tab:pairwise-results-tab}
        \footnotesize
        \setlength{\tabcolsep}{4pt}
        \begin{tabular}{@{}lcc@{}}
            \toprule
            \textbf{Pair} (winners in bold) & $\substack{\text{Winner} \\ \text{Share}}$ & \textbf{$p$ (Holm)} \\
            \midrule
            \textbf{\jankaari} vs.\ \swasthyavibhag & 100.0\% & \textbf{0.0001} \\
            \textbf{\asha} vs.\ \jankaari & 100.0\% & \textbf{0.0000} \\
            \textbf{\asha} vs.\ \swasthyavibhag & 59.1\% & 0.3938 \\
            \textbf{\ashasaheli} vs.\ \jankaari & 86.4\% & \textbf{0.0019} \\
            \textbf{\ashasaheli} vs.\ \asha & 84.2\% & \textbf{0.0057} \\
            \textbf{\ashasaheli} vs.\ \swasthyavibhag & 90.0\% & \textbf{0.0014} \\
            \bottomrule
        \end{tabular}
    \end{minipage}
\end{figure*}

The overwhelming preference for \ashasaheli was echoed in the qualitative themes, which provide rich insight into why ASHAs gravitated toward this normative framing. We unpack these themes in the section that follows.

\subsubsection{Why Narrative Identity \chidelete{Bot}\chiadd{Framing} Resonated with ASHAs}
Participants consistently reported that the \chiadd{interface of} \ashasaheli \textit{“felt like talking to another ASHA,”} fostering comfort and ease of interaction. In contrast, the \chiadd{formal, government-like tone of the interface of} \swasthyavibhag{}\chidelete{'s formal government tone} was often described as alienating. At first glance, this may seem counterintuitive: prior work shows that institutional endorsement can boost uptake, as frontline workers align with government authority \cite{Banerjee_2019, CSBC_2022}. Yet other studies caution that bureaucratic framings can invoke fear, surveillance, and distance \cite{Barnreuther2024}). Our findings align with this latter: for everyday health queries, ASHAs strongly preferred the narrative identity framing. While institutional voice may confer legitimacy for official reporting contexts, narrative-identity created a sense of psychological safety for sensitive or “everyday” concerns, making the system feel approachable and trustworthy.

Our interviews revealed three themes explaining why the \ashasaheli framing resonated with them:

\textbf{Reliance on peer support.}
Many ASHAs reported that in practice they turn first to their close peers rather than supervisors for guidance, due to both comfort and logistical constraints in reaching their supervisors and nurses. Peer networks thus serve as the first line of consultation, offering immediate and judgment-free support. By reproducing this “peer first” channel, the \chiadd{interface of} \ashasaheli resonated with their existing help-seeking practices. Rather than positioning the chatbot\chiadd{-interface} as an authority, it was interpreted as an extension of their peer network, which heightened both familiarity and trust. P14 explained:
\begin{quote}
    \textit{“If I have a question, I usually call another ASHA first. The supervisor is at the PHC (Primary Health Centre), which is often a few kilometers away... We can always call and ask, but in my anganwadi (rural child care centre) I rarely get (cellular) network, so my choice is to either find another ASHA on my way to the next anganwadi or to go to the PHC.”} -- P14.
\end{quote}

\textbf{Comfort in asking questions.}  
Participants also emphasized that the friendly tone of \ashasaheli{}'s \chiadd{interface} lowered barriers not only for routine queries but also for more sensitive or stigmatized topics. ASHAs described feeling free to raise concerns about reproductive health, menstrual irregularities, or family planning—areas where conversations with supervisors or formal channels often carried the risk of embarrassment or judgment. The \chidelete{chatbot}\chiadd{interface}’s voice was interpreted as non-judgmental and approachable, creating space for candid dialogue in situations where silence is the safer option. P43 stated:  
\begin{quote}
\textit{“I could ask openly about issues like white discharge, menstruation, or family planning. If I ask these things to a doctor or the official bot, I feel like I'm being judged. But this felt like talking to another ASHA who understands.”} -- P43.
\end{quote}  

This sense of psychological safety was critical: rather than worrying about being reprimanded or dismissed, ASHAs felt the \chidelete{bot}\chiadd{framing} provided a private, supportive channel where even “taboo” questions could be voiced without hesitation. 

\textbf{Learning through peer stories.}  
Beyond comfort, ASHAs valued how the \chiadd{interface of} \ashasaheli normalized mistakes and modeled problem-solving through stories of other ASHAs. For example, the \ashasaheli answered questions while referring to stories about other ASHAs in the village. In one of the scenarios, it said "Your fellow ASHA also struggled with this issue!" before advising about the next steps. This enabled the ASHA to identify with her peers and feel supported even when she made a mistake. P5 contrasted this with her supervisor's scolding response to an error:
\begin{quote}  
    ``\textit{Once, while entering vaccine data, I accidentally typed 1000 vials instead of 100. I didn’t know how to fix it, so I asked my supervisor and apologised for making the mistake. He scolded me in front of 15 other ASHAs, saying that if I couldn’t do such a simple thing I should just leave the job... After that I stopped asking about my mistakes. But if I can ask these questions on my phone and if it talks to me the way it [\ashasaheli] did... I would learn more quickly. It told stories of other ASHAs who had made similar mistakes. It made me feel like I could ask anything without being judged or shouted at.}'' -- P5.
\end{quote}

By signaling empathy and shared narratives, the \chiadd{interface of }\ashasaheli created a safe space for learning that supervisors and authority-based channels often failed to provide. In contrast, the \chiadd{interface of }\swasthyavibhag was consistently described as “\textit{too strict},” with its formal phrasing and directive tone discouraging casual or sensitive queries. Several ASHAs said it felt more like a training session than a conversation, which made them hesitant to use it for everyday doubts. Others emphasized that the authoritative style also carried a practical burden: responses were lengthy, bureaucratic, and read \textit{“like a government circular,” }which felt mismatched to the rapid decision-making required in the field. As P23 explained:
\begin{quote}
\textit{“When you are in the field you don’t want long, hard to read paragraphs. This [\swasthyavibhag] was too formal and heavy. It felt like it was checking me instead of helping me.”} -- P23.
\end{quote}

This difference underscores how normative framing shaped perceptions of safety: while narrative identity framing encouraged candid, stigma-free engagement, the authority framing (unintentionally) reinforced hierarchical distance.

\subsubsection{Demographic Variation in Bot Preference}

Next\chiadd{,} we examined whether age, education, or experience moderated bot preferences. No subgroup effects reached statistical significance (Figures~\ref{fig:demog-age} and ~\ref{fig:demog-exp} in Appendix B). However, descriptive analysis suggested meaningful trends. ASHAs with $\geq$15 years of tenure were \emph{more likely} to prefer peer-oriented framings (\ashasaheli or \asha): +16.2\% vs.\ less-experienced peers, $p=.18$. These patterns aligned with interview accounts. Senior ASHAs reported longstanding skepticism toward top-down directives---having “\textit{seen it all}” from the health department---and gravitated toward tools resembling a knowledgeable colleague over an official voice. P53 stated:

\begin{quote}
    ``\textit{I have been doing this job for 20 years. I know everyone in my village, they know me, and I know my training. But sometimes questions come up that are unique. During COVID, a Muslim woman was hesitant to get vaccinated because there were rumors that Hindu nurses were putting something in it causing infertility. I didn’t know what to do. The answer was not in the training handbook. In these cases I need a fellow ASHA... someone who understands the dynamics here, to tackle such misinformation. This [\ashasaheli] bot tells me what other ASHAs in my area are doing. It’s like talking to another experienced person.} -- P53.
\end{quote}

In contrast, less-experienced or younger ASHAs showed somewhat greater receptivity to the \swasthyavibhag, citing higher trust in formal authorities and less developed peer networks:

\begin{quote}
\textit{“I am new to this work, and my village is far from the subcenter so I only meet other ASHAs once a month. I don’t know all the ASHAs in my area yet. This one [\ashasaheli] said things like ‘ASHA sister in your area does this,’ but I don’t know them personally. Until I build those connections, it's easier to see what the departmental guidelines say.”} -- P2.
\end{quote}
Despite these nuances, the \chiadd{interface of} \ashasaheli consistently elicited stronger trust and willingness to engage across groups, highlighting the power of shared identity in shaping technology adoption.
Our results show that social norms strongly influenced preferences. Yet preference is not equivalent to trust. ASHAs still had to decide whether to act on the information provided---a crucial distinction given the high-stakes and sensitive contexts they confront, e.g.,
pregnancy complications, vaccine hesitancy, childhood illness. 
We therefore next examine a central question: when ASHAs consulted the chatbot\chiadd{-interface}s, did they \emph{trust} the advice—and, crucially, could they \emph{tell apart} correct from incorrect guidance? To assess this, we analyzed responses in the low-ambiguity and high-ambiguity tasks, where trust was coded as whether the ASHA indicated she would follow the chatbot’s advice (\textit{Yes/No}).

\subsection{Trust in Chatbots in Low-Ambiguity Scenarios}
\label{sec:low}
\subsubsection{Overall Trust Distribution}
We first examined whether ASHAs could distinguish between correct and incorrect advice in low-ambiguity scenarios. We found that they were significantly more likely to trust correct responses than incorrect ones ($\chi^2(1)=10.1$, $p<0.01$), following 84.7\% of correct answers but still 58.7\% of incorrect ones.
A logistic regression (Model~\ref{eq:model1}) confirmed that correct advice was nearly four times more likely to be trusted than incorrect advice (odds ratio = 3.90, 95\% CI $[1.68,\,9.06]$). 
Despite this discernment, overreliance was high---more than half of the incorrect responses were still accepted---resulting in a low \textsc{Calibrated Trust} of 26.0\% (Table~\ref{tab:trust-by-bot}). These findings suggest that even in clear-cut cases, ASHAs remained vulnerable to errors, a concerning trend in high-stakes health settings.

\begin{figure*}[]
  \centering
  \begin{minipage}[t]{0.48\linewidth}
    \centering
    \includegraphics[width=\linewidth]{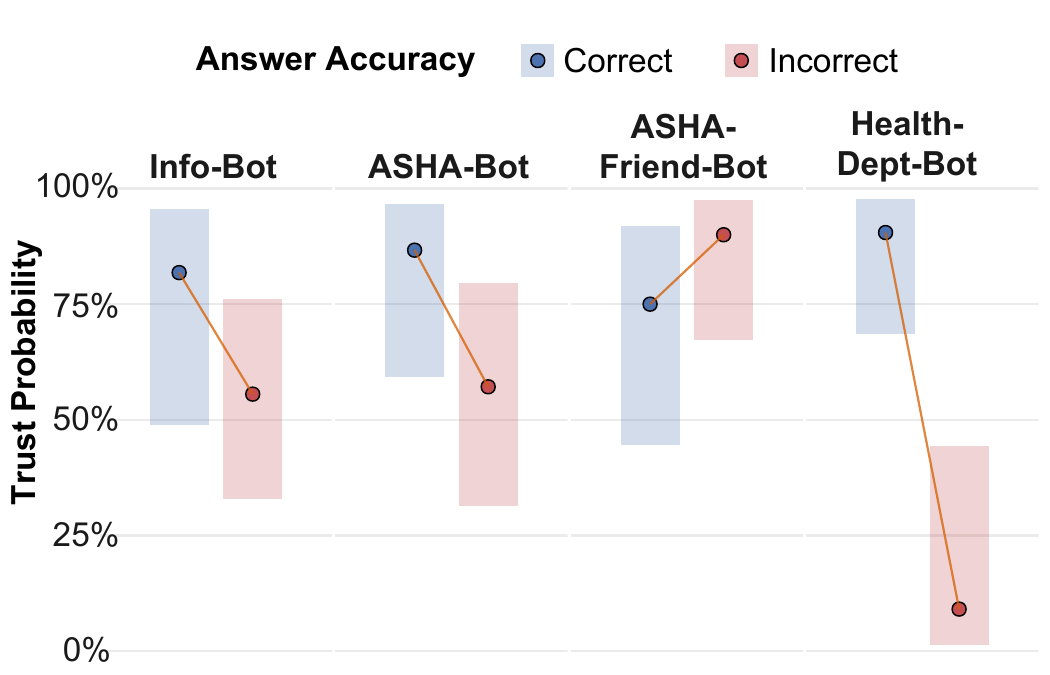}
        \caption{Trust in low-ambiguity scenarios by chatbot}
        \Description{Box plot comparing trust calibration across chatbot\chiadd{-interfaces} in low-ambiguity tasks. The figure visualizes the data from Table 6, showing that ASHA-Friend-Bot yields negative calibrated trust due to overreliance, whereas Health-Dept-Bot achieves strongly positive calibrated trust by sharply distinguishing correct from incorrect advice.}
    \label{fig:low-amb-boxes}
  \end{minipage}\hfill
  \begin{minipage}[t]{0.48\linewidth}
    \centering
    \includegraphics[width=\linewidth]{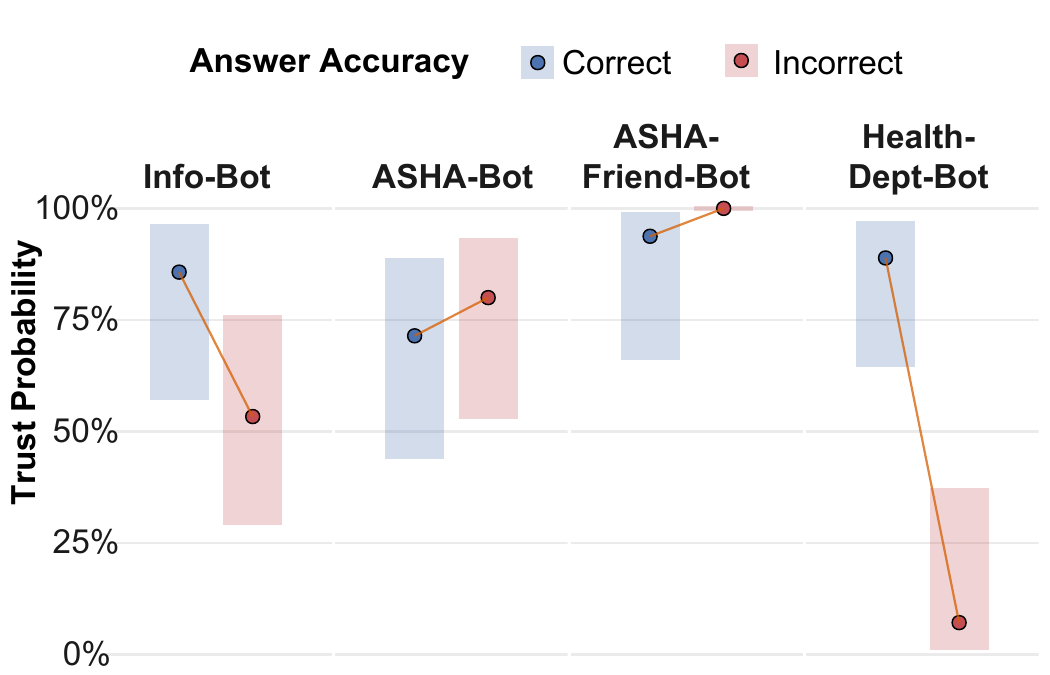}
    \caption{Trust in high-ambiguity scenarios by chatbot}
    \Description{Box plot comparing trust calibration across chatbot\chiadd{-interfaces} in high-ambiguity tasks. The figure visualizes the data from Table 6, showing that \chiadd{the interface of} ASHA-Friend-Bot again produces negative calibrated trust due to universal acceptance of advice, while \chiadd{the interface of} Health-Dept-BotBot maintains strongly positive calibrated trust by reinforcing distinctions between correct and incorrect answers.}
    \label{fig:high-amb-boxes}
  \end{minipage}
\end{figure*}

We next analyzed whether trust patterns varied across normative framings and found significant contrasts (Model 2; Figure~\ref{fig:low-amb-boxes}). \ashasaheli significantly induced \textsc{Overreliance} compared to the neutral baseline ($\gamma=1.97$, SE=0.88, $p=0.025$), with workers trusting 90\% of its incorrect advice (\textsc{Overreliance}), resulting in the lowest \textsc{Calibrated Trust} at -15.0\% ($\delta=-2.38, p=0.095$). In effect, correct and incorrect answers were accepted at nearly the same rate. By contrast, \swasthyavibhag significantly reduced \textsc{Overreliance} ($\gamma=-2.53$, SE=1.16, $p=0.030$), with 90.5\% of its correct advice being trusted, while only 9.1\% of its incorrect answers were followed. This produced the strongest \textsc{Calibrated Trust} of 81.4\% ($\delta=3.27, p=0.040$), approaching the ideal pattern of accepting correct advice but rejecting incorrect advice. \asha was very similar to the baseline, with an \textsc{Overall Trust} of 72.4\% and \textsc{Overreliance} of 57.1\%

Together, these findings highlight a trade-off: narrative identity framings (\ashasaheli) built affinity but encouraged uncritical acceptance, whereas authoritative framings (\swasthyavibhag), though less preferred, acted as safeguards against error.

\mobilehciadd{For completeness, we also examined the inverse error---underreliance, i.e., rejecting correct advice. Across framings, ASHAs rejected correct advice most often when it came from ASHA-Friend-Bot (25.0\%) and least often from Health-Dept-Bot (9.5\%), with Info-Bot (18.2\%) and ASHA-Bot (13.3\%) in between. These differences were not statistically significant ($\chi^2(3) = 1.53$, $p = .68$), and are treated as descriptive. We discuss the rationale for focusing on overreliance in Section~\ref{sec:discussion}.} 

\subsubsection{Reasons for \textsc{Overreliance} in Narrative Identity Bot and \textsc{Calibrated Trust} in Authority Bot}
Analysis of our qualitative data revealed two distinct themes that explain the contrasting trust patterns we observed. 

\textbf{Peer-like reassurance encouraged fast acceptance.}
ASHAs described the \chiadd{interface of} \ashasaheli as offering confirmation rather than new information. In low-ambiguity questions (e.g., vaccination schedule), where ASHAs already knew the answer, its friendly tone acted like a nod of agreement from a colleague, accelerating compliance without reflection. As P14 put it:
\begin{quote}
\textit{``This was something I already do every day. When it [\ashasaheli] confirmed it in a friendly way, I didn’t stop to think. It was just a quick yes.’’} -- P14.
\end{quote}
Here, affinity and familiarity blurred the boundary between peer talk and factual verification, reducing critical scrutiny.

\begin{table*}[t]
\centering
\small
\setlength{\tabcolsep}{4pt}
\caption{Trust metrics across chatbots in low- and high-ambiguity tasks.}
\Description{This table reports four trust metrics--Overall Trust, Appropriate Trust, Overreliance, and Calibrated Trust--for each chatbot framing under low- and high-ambiguity tasks. In low-ambiguity settings, ASHAs showed high Overall Trust across all bots, but ASHA-Friend-Bot exhibited the strongest Overreliance (90\%), giving a negative Calibrated Trust of -15.0\%. By contrast, Health-Dept-Bot achieved the best Calibrated Trust (81.4\%) by sharply distinguishing correct from incorrect advice. Under high-ambiguity, Overall Trust remained high, but calibration diverged further: ASHA-Friend-Bot accepted all incorrect answers (Calibrated Trust -6.2\%), while Health-Dept-Bot again maintained strong calibration (81.8\%).}
\label{tab:trust-by-bot}
\begin{tabular}{@{}l cccc cccc@{}}
\toprule
\multirow{2}{*}{\textbf{Chatbot}} &
\multicolumn{4}{c}{\textbf{Low-ambiguity task}} &
\multicolumn{4}{c}{\textbf{High-ambiguity task}} \\
\cmidrule(lr){2-5}\cmidrule(lr){6-9}
 &
\textbf{$\substack{\text{Overall} \\ \text{Trust}}$} &
\textbf{$\substack{\text{Appropriate} \\ \text{Trust}}$} &
\textbf{$\substack{\text{Over-} \\ \text{reliance}}$} &
\textbf{$\substack{\text{Calibrated} \\ \text{Trust}}$} &
\textbf{$\substack{\text{Overall} \\ \text{Trust}}$} &
\textbf{$\substack{\text{Appropriate} \\ \text{Trust}}$} &
\textbf{$\substack{\text{Over-} \\ \text{reliance}}$} &
\textbf{$\substack{\text{Calibrated} \\ \text{Trust}}$} \\
\midrule
\jankaari        & 65.5\% & 81.8\% & 55.6\% & 26.3\%    & 69.0\% & 85.7\% & 53.3\%  & 32.4\%   \\
\asha            & 72.4\% & 86.7\% & 57.1\% & 29.5\%    & 75.9\% & 71.4\% & 80.0\%  & $-8.6$\% \\
\ashasaheli      & 84.4\% & 75.0\% & 90.0\% & $-15.0$\% & 96.9\% & 93.8\% & 100.0\% & $-6.2$\% \\
\swasthyavibhag  & 62.5\% & 90.5\% & 9.1\%  & 81.4\%    & 53.1\% & 88.9\% & 7.1\%   & 81.8\%   \\
\bottomrule
\end{tabular}
\end{table*}


\textbf{Authoritative tone disrupted routine and prompted reflection.}
By contrast, the \chiadd{interface of} \swasthyavibhag increased \textsc{Calibrated Trust} by slowing ASHAs down. 
Its formal, bureaucratic style stood out against other conversational norms, even for simple day-to-day questions. Participants reported that this unfamiliarity made them pause and compare the advice against their own knowledge. P27 stated:
\begin{quote}
\textit{``I knew the right answer already, but when it came in that official style I slowed down and compared. That’s how I saw the mistake.’’} -- P27.
\end{quote}
Rather than reinforcing existing beliefs, the authoritative framing interrupted routine acceptance, creating space for reflection and error detection.

\subsubsection{Demographic Variation in Trust by Bot}  
Finally, we examined whether demographics shaped these outcomes (Model \ref{eq:model3}). Adding age, education, and experience did not alter the main results, and none were statistically significant. However, descriptive patterns suggest that more experienced and educated ASHAs were less likely to follow incorrect advice (see Figure \ref{fig:wrong-edu-overall} in Appendix). 
As one veteran noted: “\textit{There is nothing in these tasks that I haven’t seen first-hand... If a bot—or even an ANM—gave the wrong instruction here, I would catch it and not follow it}” (P21).
These reflections help interpret the negative (though non-significant) coefficients,
they represent embodied familiarity that can counteract \chideletenew{blind trust}\chiaddnew{overreliance} in routine matters. 


\subsection{Trust in Chatbots in High-Ambiguity Scenarios} 
\label{sec:high}
We now turn to high-ambiguity scenarios, where the correct answer is unclear or contested, and the central challenge is navigating uncertainty.
While low-ambiguity tasks revealed whether ASHAs could distinguish correct from incorrect advice in routine matters, high-ambiguity tasks examined whether they could still exercise such discernment when certainty itself was in question.

\subsubsection{Overall Trust Distribution} 
Logistic regression (Model \ref{eq:model1}) showed that ASHAs were nearly five times more likely to follow correct than incorrect advice ($\beta=1.60$, SE=0.52, $p=.002$; odds ratio = 4.94, 95\% CI [1.8, 13.4]). 
The gap between following correct and incorrect advice narrowed compared to low-ambiguity tasks: \textsc{Overall Trust} remained high at 73.8\%, but pooled \textsc{Calibrated Trust} dropped from 26.0\% to  23.8\% (Table~\ref{tab:trust-by-bot}). 
\textsc{Overreliance} also increased, with ASHAs accepting 61.7\% of incorrect answers. 
Together, these results show that high-ambiguity scenarios weakened ASHAs’ ability to tell correct advice from incorrect, making them more likely to follow wrong recommendations.

These effects varied significantly across normative  framing (Model~\ref{eq:model2}; Figure~\ref{fig:high-amb-boxes}). \ashasaheli significantly increased \textsc{Overreliance} compared to the neutral baseline ($\gamma=1.97$, SE=0.88, $p=0.025$), with near-universal compliance: 96.9\% of its correct answers and 100\% of its incorrect ones were followed, a jump from 90\% in low-ambiguity scenarios. This led to minimal \textsc{Calibrated Trust} of -6.2\% ($\delta=-2.38$, $p=0.095$), which is better than what we found in low ambiguity scenarios (of -15.0 \%). However, the fact that 100\% of the incorrect advice was also trusted indicates \chideletenew{blind acceptance}\chiaddnew{overreliance} over any advice given by this bot\chiadd{'s interface}, and not better discernment. In other words, participants appeared to follow this bot’s advice uncritically, regardless of accuracy.

By contrast, \chiadd{the interface of }\swasthyavibhag continued to act as a safeguard. It significantly reduced \textsc{Overreliance} ($\gamma=-2.53$, SE=1.16, $p=0.030$), with only 7.1\% of its wrong answers trusted (down from 9.1\% in low-ambiguity), while 88.9\% of its correct answers were followed. This produced the strongest \textsc{Calibrated Trust} of 81.7\% ($\delta=3.27$, $p=0.040$). Here, the \textsc{Calibrated Trust} increased slightly from what we observed in low-ambiguity scenarios. These results show that authority framings maintained their protective effect even under uncertainty, helping ASHAs resist following incorrect advice. Finally, \chiadd{the interface of} \asha did not differ significantly from baseline, but its calibration worsened under ambiguity ($p=0.045$), suggesting that what felt like reassuring advice from other ASHAs in clear cases became a liability when answers were uncertain. 

\mobilehciadd{For underreliance, the pattern reversed and was not statistically significant ($\chi^2(3) = 3.28$, $p = .35$). Under high ambiguity, ASHAs rejected correct advice most often from ASHA-Bot (28.6\%) and least often from ASHA-Friend-Bot (6.2\%), with Info-Bot (14.3\%) and Health-Dept-Bot (11.1\%) in between.}

\subsubsection{Why High Ambiguity Impacted Calibrated Trust}
Our qualitative interviews surfaced two interlocking dynamics that explain why, under high-ambiguity, overreliance reached 100\% for \chiadd{the interface of} \ashasaheli and increased even for \chiadd{the interface of} \swasthyavibhag compared to low-ambiguity scenarios.

\textbf{Social proof substituted for missing knowledge}.
In high-ambiguity scenarios, ASHAs admitted they were unsure of the \textit{``correct''} practice, hence the peer narrative story itself became treated as evidence. Advice presented as what \textit{``another ASHA sister''} had done was inferred to be valid as it reflected lived village realities. P2 explained:
\begin{quote}
\textit{``I wasn’t sure what to do, but if it says another ASHA here does it, then I believe it must be right. She knows what happens in our kind of village, so I trusted it.’’} -- P2
\end{quote}
Here, narrative identity from a peer filled the gap left by unclear or evolving protocols, effectively substituting social proof for technical knowledge.

\textbf{Comfort muted doubt}.
Ambiguity usually provokes hesitation, but the warm and familiar tone of the narrative bot made advice feel safe to accept without scrutiny. P33 reflected:
\begin{quote}
\textit{``Actually I wasn’t sure if this was right. The bot told me that if the MR vaccine is missed at nine months, it should not be given later. I had never heard of that before, so I was confused. But then it also said that another ASHA from my PHC had advised the same, so I didn’t think twice before agreeing. 
I thought maybe it 
was some new update I hadn’t been trained on.’’} -- P33
\end{quote}
Here, comfort amplified by uncertainty muted critical attention, especially among less-experienced ASHAs.
Peer endorsement was interpreted as a possible signal of updated guidance, making incorrect responses more likely to slip through,
further reducing the impulse to double-check. 
By contrast, participants often described the \swasthyavibhag{}’s distance—both in language and tone—as prompting a different kind of reasoning. 
Its formality reactivated memories of official training, guidelines, and “best practices,” even when these clashed with field realities. P60 shared: 
\begin{quote}
    \textit{“The way the chatbot spoke, it reminded me of my training days...
    But when you work in the field you realize not everything that you are taught [in training] is possible.
    E.g., when it comes to institutional delivery, the chatbot advised to push for it, even if the family is not interested [in institutional delivery]. But sometimes you can't push, because the family is very aggressive about their choice and don't want my involvement...
    But it [bot] spoke so formally and like a government person... it said it is my duty as an ASHA... it is my responsibility... it reminded me that it is a best practice to try even if we expect that the family will not change their mind.”} -- P60
\end{quote}

These findings show that normative framing effects shifted with varying task ambiguity. In low-ambiguity scenarios, narrative framing sped up acceptance and felt like friendly confirmation, while authority framing slowed ASHAs down and prompted careful checking. In contrast, in high-ambiguity scenarios, narrative framing became a substitute for missing knowledge, encouraging \chideletenew{blind acceptance}\chiaddnew{overreliance}. 


\subsubsection{Demographic Variation in Trust}  
Finally, we examined whether demographics moderated these effects (Model \ref{eq:model3}). While no coefficients were statistically significant, descriptive trends were consistent. Less-experienced ASHAs were the most trusting of \ashasaheli, following 97.4\% of its advice. By contrast, more experienced workers were about 15\% less trusting overall, reflecting greater caution toward narrative identity framings. This echoes qualitative accounts that experience not only deepened familiarity with protocols but also cultivated a healthy skepticism toward both digital tools and top-down instructions.


\section{Discussion}
\label{sec:discussion}
Our study shows that normative framings in health chatbots deeply impact both \emph{preference} and \emph{trust}. Here, we discuss three implications of our findings on HCI design and research.

\subsection{Culture as a Double-Edged Lever}
Calls to ``design with culture'' \cite{Marcus2000, BarberBadre1998, Sun2012, VanBoeijen2020} in HCI often assume that cultural alignment is uniformly beneficial. Our results complicate that claim by focusing on one specific cultural lever---\emph{social norms}. We find that social norms are powerful because they reshape judgments about legitimacy and appropriateness, not just aesthetics. Our results align with prior work showing that narrative identity markers in conversational agents (e.g., anthropomorphic language, human-like representation, communicative agency) shift perceived credibility, warmth, and willingness to comply \cite{Araujo2018,Park2023,Seeger2021}. 
On the other hand, the literature on injunctive norms further shows that authoritative sources can heighten perceived obligation \cite{Hofstede2001,HovlandWeiss1951}. Where we extend the literature is in demonstrating a reliability trade-off: the same cues that elevate \emph{preference} can degrade \emph{calibrated trust}. Narrative identity framings are highly engaging yet prone to overreliance, whereas injunctive authority framings dampen appeal but improve error discernment. Our evidence thus cautions against blind maximization of cultural cues (in our case, of social norms) and the universal directive to embed such cues for adaptability; instead, it supports treating cultural cues as a tunable design lever with measurable preference and safety trade-offs. 

\chiaddnew{These findings also help explain why prior broadcast-based norm interventions  have yielded largely null effects~\cite{Papakonstantinou2025}. In SMS  or poster campaigns, normative messages arrive without relational context—they cannot embody a peer identity, respond to individual concerns, or build rapport over conversational turns. Our results suggest that normative cues may require relational grounding to influence behavior: the \textit{ASHA-Friend-Bot's} peer framing was powerful precisely because it felt like a conversation with a colleague, not a broadcast announcement. This implies that conversational AI can unlock normative influence in ways that broadcast media cannot, but 
that this heightened influence carries corresponding risks of overreliance when the system provides incorrect guidance.}

For chatbot design, this implies that instead of hard-coding a specific normative framing, the bot should flexibly adapt its responses based on the severity and sensitivity of the user's query, the bot's confidence in its answer, the trajectory of the conversation, and the experience level of the user. 
In practice, one way to achieve it is by tuning the normative framing lever using observable signals from the system, such as: \textit{interaction mode}—is the user exploring the bot (many short, unrelated, open-ended questions) or moving towards committing to an action (many focused questions about the same issue)?; and \textit{who is asking}—a new vs. experienced user.
In exploratory interactions or with newer users, use narrative identity framings to encourage curiosity and support learning. When the exchange shifts toward commitment (e.g., questions regarding supplies, doses, or issuing referrals), the bot should switch to an authoritative voice and show the relevant guideline citation upfront.

\subsection{Ambiguity as a Stress Test}
We found that normative framing mattered most when ``the right advice'' collided with what was actually doable on the ground. In routine work, a narrative identity voice felt natural and was readily accepted. In contested situations, the same voice often sped people toward agreement even when they felt unsure. An injunctive authority voice showed the mirror image: it could feel rigid and distant in low-ambiguity tasks, but under uncertainty, these characteristics acted like a guardrail by pulling training and guidelines back into view and prompting ASHAs to return to ``what should be done''. In short, ambiguity revealed when cultural cues steady judgment and when they destabilize it. 

\chiadd{Our findings on ASHAs overrelying on chatbots are consistent with HCI literature on overreliance in AI-powered decision support tools \cite{Bucinca2021}, as well as prior work showing community health workers \chidelete{place near-blind trust}\chiaddnew{show overreliance} on AI outputs, attributing inherent accuracy and authority to "the machine" \cite{Okolo2021, Okolo2024}. Where we extend the literature is in demarcating the areas where this \chideletenew{blind acceptance}\chiaddnew{overreliance} takes full force---in spaces of high ambiguity.}
Prior research shows that ambiguity changes how people weigh system advice and increases reliance on perceived expertise \cite{Zhang2020,Deutsch1955,Baron1996}. Why, then, did ASHAs overrely so heavily on the \ashasaheli{}—a source not framed as authoritative—when things were unclear? Our findings uncover a salient type of ambiguity in healthcare contexts (particularly prevalent in low- and middle-income countries): ambiguity of feasibility. In our high-ambiguity scenarios, there was a protocol-driven correct answer that ASHAs were trained on, but it conflicted with on-the-ground constraints (e.g., family pressures, travel distance, resource unavailability, local contextual considerations). In these situations, their close peers functioned as \emph{situated experts}: they knew which workarounds were locally feasible, which steps families would accept, and which referrals would actually go through. When protocol and practice were misaligned, the narrative identity framing gained legitimacy because it better matched what could realistically be done. Our findings thus extend the literature by formulating a distinction between \emph{feasibility-driven} and \emph{informational} ambiguity---showing that who is considered an authority can shift from formal institutions to experienced peers depending on the kind of ambiguous situation.

\chideletenew{ASHAs' willingness to follow the peer-framed chatbot thus reflects not uncritical acceptance but a transfer of socially constructed trust from experienced colleagues to a system framed as being one of them.}\chiaddnew{This raises a critical question: does this willingness to follow the peer-framed chatbot reflect genuine irrationality, or a rational transfer of trust?} In their day-to-day practice, trusting peer-endorsed advice is a learned heuristic that typically serves \chideletenew{them}\chiaddnew{ASHAs} well: peer knowledge often captures on-the-ground feasibility and social constraints more accurately than distant protocols. \chideletenew{Although we describe this behavior with the terminology of "overreliance" or "blind acceptance," such framing risks obscuring the rationality of ASHAs' responses. What appears as blind trust may therefore be a form of situated trust, transferred from real peer networks to the chatbot's constructed peer voice. This reframing highlights that the design problem is not that ASHAs trusted incorrectly, but that the chatbot effectively exploited a functional heuristic by pairing incorrect guidance with credible social markers. The implication for HCI and AI design is thus not to "debias" frontline workers, but to ensure that systems do not misuse culturally grounded trust cues or mimic peer legitimacy without accountability. However, even with this critical reframing, it is important to note that situated trust or blind acceptance on peer networks in health scenarios pose risks for safety and accuracy when the information source--whether human or bot--may be wrong.} \chiaddnew{This is why, we argue that while the cognitive mechanism is \textit{situated trust}---a functional heuristic favoring peer-endorsed knowledge—the behavioral outcome in a safety-critical AI context is overreliance. By "overreliance", we do not imply a lack of competence; rather, we describe a pattern where ASHAs defer to system outputs without verification or sensemaking at the point of decision. This distinguishes their behavior from "calibrated trust," where users actively negotiate or condition their reliance. The danger of conversational AI lies in its ability to exploit this situated trust by mimicking peer markers without inheriting the interpersonal accountability that regulates real-world networks. The implication for HCI is thus not to "debias" frontline workers, but to recognize that when situated trust is transferred to an unaccountable system, it becomes dangerous overreliance. Design must therefore focus on ensuring systems do not mimic peer legitimacy without accompanying safeguards.}

\chiaddnew{These insights point toward concrete design directions.} From a design perspective, ambiguity should be treated not only as missing information but also as limited feasibility.
When cues of infeasibility arise (e.g., repeated mentions of distance, or family refusal), health chatbots should acknowledge these constraints explicitly and pair protocol guidance with narrative-informed, locally workable next steps (e.g., follow-ups with families or acceptable substitutes). Thus, designing \emph{dynamic norm systems} that adjust style--and even combine normative framings based on context--is a promising direction.
Hybrid framings---messages that combine normative identity with explicit reference to guidelines—may balance user resonance with accountability. For example, a chatbot\chiadd{-interface} might say, \textit{‘Radhika, an ASHA in your area, continued counseling for institutional delivery when families hesitated—this is also what Health Department guidelines say.’} In this, the narrative identity norm precedes the injunctive authority in the sentence. The positioning of the norm could also change based on user profile. Empirical evaluation of such adaptive and hybrid approaches, ideally in real-world deployments over extended periods, will be critical for assessing whether norm-based chatbots can sustain calibrated trust at scale.

\subsection{From Trust to Calibrated Trust}
A common goal in HCI is to “build trust” in AI, especially in health contexts~\cite{Asan2020, Bickmore2005, cataractbot2025}. Our findings suggest this framing is insufficient. What matters is \emph{calibrated trust}: workers should follow advice when it is correct and resist or override it when it is incorrect. Prior work has made a similar point—trust should align with actual performance and context—and also shown that surface-level trust boosts (e.g., human-like cues or smooth UX) can increase reliance even when the system is wrong \cite{Kocielnik2019,Liao2022,Bansal2021}. Showing case-by-case reliability signals (like model confidence) can help, but these alone are not enough; calibrated trust must be \emph{designed}, \emph{measured}, and \emph{optimized}, rather than inferred indirectly from satisfaction or preference scores \cite{Zhang2020,Bansal2021}. 

Our findings add a simple but important twist: \emph{social norms} shape what feels trustworthy. Normative framings can make people say “yes, I trust” even when the advice is wrong (as with overreliance on \ashasaheli), or make them hesitate even when the advice is right. To support calibrated trust, reliability signals should be paired with dynamic norm cues. For instance, a chatbot might use assuring language in high certainty cases such as tablet doses and vaccination schedules (‘you should,’ ‘you must,’ ‘you need to’) but adopt more cautious phrasing in contested scenarios (‘best practice is usually…,’ ‘it is recommended…’). This linguistic modulation should work in tandem with dynamic switching of normative framings based on context and user profile. Prior evidence~\cite{Bucinca2021,Zhang2020,Shuai2024} shows that prompts like ``think twice'' and ``are you sure?'' reduce overreliance by creating a moment of distance which promotes discernment. Our findings show that injunctive authority works in a similar way, as formal framing forces ASHAs to pause and rethink. Thus, pairing such framings in chatbot communication can promote calibrated trust. 

\chiadd{Designing for calibrated trust then also requires acknowledging that these systems will occasionally fail. Instead of hiding that fallibility, chatbots should surface uncertainty and guide workers toward safe next steps. This can be done by: (a) explicitly showing low confidence (“I’m not certain about this”), (b) briefly explaining the basis for its advice (“This answer draws on Health Department guidelines”, "This reflects common practices reported by ASHAs in your area"), and (c) enabling workers to easily report or correct inaccurate responses ("Report this answer as incorrect"). Making uncertainty visible and actionable not only mitigates potential harm but also fosters more sustained engagement and trust, encouraging ASHAs to rely on the chatbot for routine guidance while being directed to other trusted sources of knowledge, such as supervisors, training manuals, or peer networks, when the system’s confidence is low.}

Methodologically, \chidelete{this means that evaluations of }\chiadd{this pushes us to rethink evaluation of such systems.} AI-powered health chatbots should report \emph{calibration} directly—alongside trust, usability, or preference. The strength of these technologies should hence be judged by whether they \emph{improve calibration} and downstream outcomes, not merely whether they boost trust. Through this lens, our work reframes “cultural alignment” in healthcare: the aim is not to maximize resonance, but to \emph{calibrate} it—dialing norm cues up or down with model confidence, task criticality, and user state so that reliance stays in check \cite{Bansal2021,Liao2022,Zhang2020}.

\subsection{\mobilehciadd{The Asymmetry Between Over- and Underreliance}}
\mobilehciadd{A natural question is whether the same theoretical machinery we used for overreliance also applies to its counterpart, underreliance. While the descriptive patterns for underreliance reported in Section~\ref{sec:findings} are suggestive, they are not statistically significant. More importantly, our paper focuses on overreliance for two reasons.} 

\mobilehciadd{First, the two errors reflect different constructs. Overreliance concerns failures of the AI system: cases in which the system produces an incorrect response and the user fails to detect it~\cite{parasuraman1997, lee2004}. Underreliance, in contrast, reflects the user's relationship to her own expertise, where she trusts her training and her peers over the system output. The norm theories we draw on (descriptive, narrative identity, injunctive) explain compliance with social cues, making them more directly applicable to overreliance. Why people resist to AI systems is better addressed through adjacent work on reactance~\cite{brehm1966, steindl2015}, algorithm aversion~\cite{dietvorst2015, logg2019}, and professional identity protection~\cite{beane2019, christin2017}.} 

\mobilehciadd{Second, the two errors carry unequal consequences. If an ASHA rejects a correct answer, she falls back to the existing standard of care--her training, supervisors, and peers; the system has at worst failed to add value. If she acts on an incorrect answer, she may give the wrong dose or push a family in the wrong direction, and that error compounds into patient care. In such settings, being conservative is usually safe and being confidently wrong is not; thus, overreliance is the more consequential failure mode and the one worth designing against first.}

\subsection{Limitations and Future Work}

Our study has limitations that suggest concrete directions for future research. First, our sample was restricted to 61 ASHAs in rural Rajasthan, and their cultural and professional context likely shaped how framings were perceived. Replication with more diverse groups—such as urban health workers or users in more individualist contexts—would clarify how cultural setting moderates normative effects.

\chiadd{Second, we used single-turn static chatbot mocks rather than a live, interactive system. While this design strengthens our study by isolating framing effects---consistent with prior HCI work using static mockups for chatbot evaluation~\cite{Cox_2022, Yun_2025, Metzger_2024}---it abstracts away the dynamics of real conversations. Multi-turn long-term interactions may amplify normative effects through increased engagement and social presence, or attenuate them as users recalibrate trust based on repeated experience. Future field deployments are therefore necessary to understand how these effects evolve in practice.} \mobilehciadd{We hypothesize three directions in which these effects may evolve in live deployments: (i) narrative framings may strengthen through sustained rapport or weaken with familiarity as the ``peer voice'' begins to feel scripted; (ii) authority framings may routinize and lose their pause-and-reflect effect, or accrue legitimacy as institutional grounding is repeatedly reinforced; and (iii) repeated exposure to errors may drive trust recalibration, especially if the system surfaces uncertainty or supports user-led error reporting.}

\chidelete{Second }\chiadd{Third}, our scenario-based design captured only short-term interactions. While useful for isolating framing effects, it cannot reveal how trust develops across repeated use. Longitudinal into-the-wild deployments are therefore a key avenue. Future work should examine whether the appeal of narrative framings wanes with familiarity, whether repeated exposure to errors prompts recalibration, and how authority framings play out in everyday practice. Larger and more heterogeneous samples would also help uncover demographic trends (e.g., whether younger or less experienced workers are more vulnerable to overreliance).

\chidelete{Third }\chiadd{Fourth}, our study focused specifically on health contexts. It remains an open question whether the same normative framing effects extend to other high-stakes domains such as education, law, or agriculture, where social expectations and authority cues may differ. Exploring these cross-domain dynamics would clarify the generality of our findings and help guide the design of culturally adaptive chatbots more broadly.

\chiadd{Lastly, our normative framings were derived from established social psychology categories widely used in health communication and behavioral design research. These framings were applied top-down rather than co-developed with ASHAs. While this ensures theoretical grounding and comparability with prior work, it risks imposing external conceptual structures onto local ways of reasoning about trust and advice. ASHAs may draw on framings that do not align neatly with academic taxonomies, for instance, distinctions based on caste, kinship, or institutional histories that shape whose voice carries weight. Future work should use qualitative methods~\cite{gtm} to examine how ASHAs describe and interpret social influence, potentially validating, extending, or challenging the categories we employed.}


\section{Conclusion}
This paper investigates how social norm framings in health chatbots shape frontline workers' \emph{preference} and \emph{trust}. We designed four otherwise-identical chatbots that differed only in social norms (descriptive, narrative identity, injunctive authority, and neutral) and evaluated them with ASHAs in rural India. 
Our findings reveal a tradeoff: narrative identity norm maximizes comfort and preference but increases overreliance, while injunctive authority is less preferred yet promotes calibrated trust. These effects intensify in scenarios where formal protocol guidance collides with on-the-ground feasibility. We translate these insights into design recommendations for AI-powered health chatbots: adopt \emph{dynamic norm framings} that switch by scenario; incorporate \emph{feasibility-aware advice} when protocol and on-ground realities diverge, and use \textit{hybrid norms} that combine narrative identity with injunctive authority to balance comfort with safety. Finally, we argue that evaluation practices should go beyond trust, usability, or satisfaction alone. Reporting \emph{calibrated trust}---follow when correct and resist when incorrect---should become a standard for safety in health chatbots. 
More broadly, our work reframes “cultural alignment” in HCI: cultural cues should not simply be maximized, but calibrated, since their uncritical amplification can harm as much as help in risk-sensitive domains like healthcare.

\section{Acknowledgements}
We thank the ASHA workers who generously shared their time, insights, and experiences, without whom this research would not have been possible. We are deeply grateful to our partner NGO, \textit{Khushi Baby}, for their critical role in planning and coordinating government permissions for fieldwork. Their support was instrumental in ensuring smooth implementation of the study.

\bibliographystyle{ACM-Reference-Format}
\bibliography{references}

\clearpage
\appendix
\setcounter{figure}{0}
\renewcommand{\thefigure}{A\arabic{figure}}
\setcounter{table}{0}
\renewcommand{\thetable}{A\arabic{table}}
\onecolumn

\section*{Appendix}
\label{appendix}

\subsection*{Appendix A: Power Calculation Details}
\label{appendix:power}
\chiaddnew{The effective observation counts were computed using the standard cluster design-effect adjustment to account for repeated measures from the same ASHA, calculated as:
\begin{equation}
N_{\text{eff}} = \frac{N_{\text{raw}}}{1 + (n - 1)\rho}
\end{equation}
where $n$ is the number of tasks completed per ASHA and $\rho$ is the intra-class correlation (ICC) across responses. We estimated plausible ICC values (0.10--0.20) using random-intercept models at the ASHA level. Power simulations~\cite{duflo2006toolkit} confirmed that this design achieved approximately 80\% power to detect main framing effects of 12--15 percentage points at $\alpha{=}0.05$. These values are typical for clustered behavioral experiments and were sufficient to detect main framing effects (e.g., whether one chatbot framing was systematically preferred, whether ASHAs calibrated trust by following correct more often than incorrect advice), though not subtle subgroup interactions (e.g., framing $\times$ ambiguity $\times$ education).}

\subsection*{Appendix B: Statistical Models}

We report three logistic regression models used to estimate trust outcomes. In all models, the dependent variable is whether worker $i$ in task $j$ \textit{trusted} the chatbot's advice ($p_{ij}$). Results are reported as odds ratios with cluster-robust standard errors.  

\paragraph{Model 1: Overall accuracy effect.}

\begin{equation}
\label{eq:model1}
\operatorname{logit}(p_{ij}) = \alpha + \beta \,\textsc{Accuracy}_{ij}
\end{equation}

\noindent

This baseline model estimates whether ASHAs were more likely to follow correct versus incorrect advice, regardless of chatbot framing.  
\begin{itemize}
  \item $p_{ij}$: probability that worker $i$ follows advice in task $j$.  
  \item $\alpha$: intercept (baseline log-odds of following wrong advice).  
  \item $\beta$: effect of accuracy (increase in log-odds of following when advice is correct).  
  \item $\textsc{Accuracy}_{ij}$: indicator variable = 1 if advice is correct, 0 if incorrect.  
\end{itemize}

\paragraph{Model 2: Chatbot differences.}
\begin{equation}
\label{eq:model2}
\operatorname{logit}(p_{ij})
= \alpha + \beta\,\textsc{Accuracy}_{ij}
+ \sum_{k\neq b_0} \gamma_k\,\mathbf{1}\{\textsc{Bot}=k\}
+ \sum_{k\neq b_0} \delta_k \,(\textsc{Accuracy}_{ij}\times \mathbf{1}\{\textsc{Bot}=k\})
\end{equation}

\noindent
This model adds chatbot framing to test whether different bots shift levels of trust or calibration relative to the neutral baseline.  
\begin{itemize}
  \item $\gamma_k$: main effect of chatbot $k$ (difference in \textit{Overall Trust} relative to baseline, when advice is wrong).  
  \item $\delta_k$: interaction between chatbot $k$ and accuracy (difference in \textit{Calibrated Trust} relative to baseline).  
  \item $\mathbf{1}\{\textsc{Bot}=k\}$: indicator for chatbot persona $k$.  
  \item $b_0$: reference bot, the neutral \jankaari.  
\end{itemize}

\paragraph{Model 3: Demographic moderation.}
\begin{equation}
\label{eq:model3}
\begin{aligned}
\operatorname{logit}(p_{ij})
={}& \alpha + \beta\,\textsc{Accuracy}_{ij}
   + \sum_{k\neq b_0} \gamma_k\,\mathbf{1}\{\textsc{Bot}=k\} \\
   &+ \sum_{k\neq b_0} \delta_k\,(\textsc{Accuracy}_{ij}\times \mathbf{1}\{\textsc{Bot}=k\}) \\
   &+ \boldsymbol{\theta}^\top \mathbf{D}_i
   + \boldsymbol{\phi}^\top (\textsc{Accuracy}_{ij}\times \mathbf{D}_i)
\end{aligned}
\end{equation}
\noindent
This full model adds worker demographics to test whether age, education, or experience moderate trust or calibration.  
\begin{itemize}
  \item $\mathbf{D}_i = (\textsc{Age}_i,\ \textsc{Education}_i,\ \textsc{Experience}_i)$: demographic covariates for worker $i$.  
  \item $\boldsymbol{\theta}$: coefficients for main effects of demographics (baseline trust differences).  
  \item $\boldsymbol{\phi}$: coefficients for interactions between demographics and accuracy (differences in \textit{Calibrated Trust} by subgroup).  
\end{itemize}

\medskip
\noindent
In all models, standard errors are clustered at the participant level to account for repeated measures.

\clearpage

\subsection*{Appendix C: Demographic Variation in Bot Preference}
\begin{figure*}[h]
  \centering
  \subfloat[Age (years)]{\includegraphics[width=0.45\linewidth]{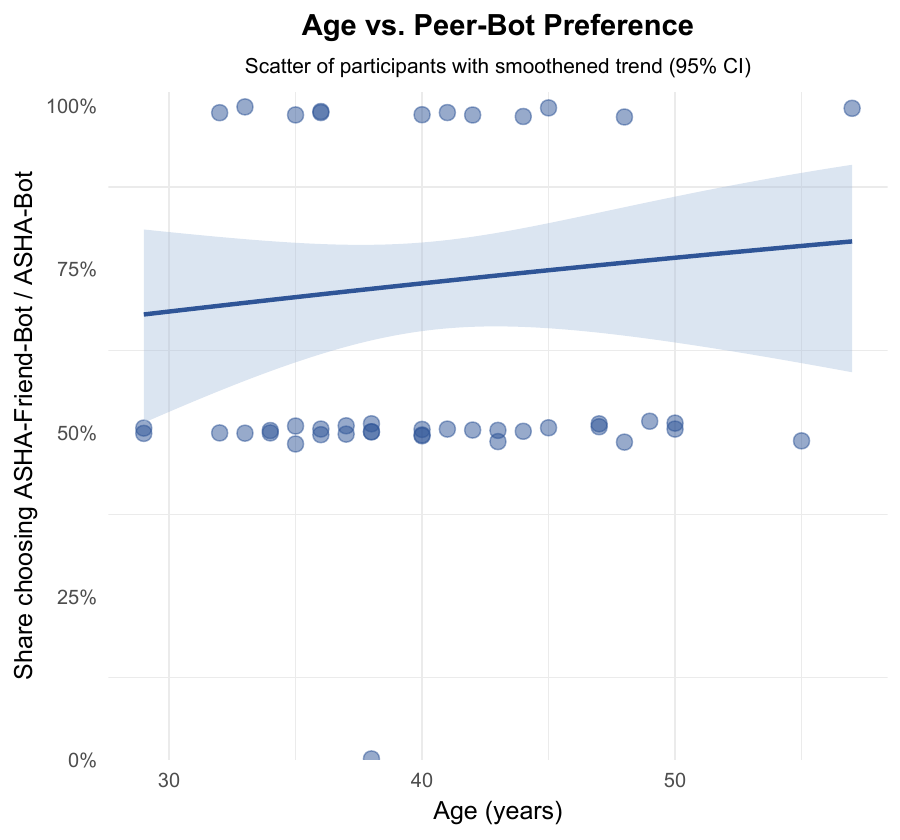}\label{fig:demog-age}}
  \hfill
  \subfloat[Experience (years)]{\includegraphics[width=0.45\linewidth]{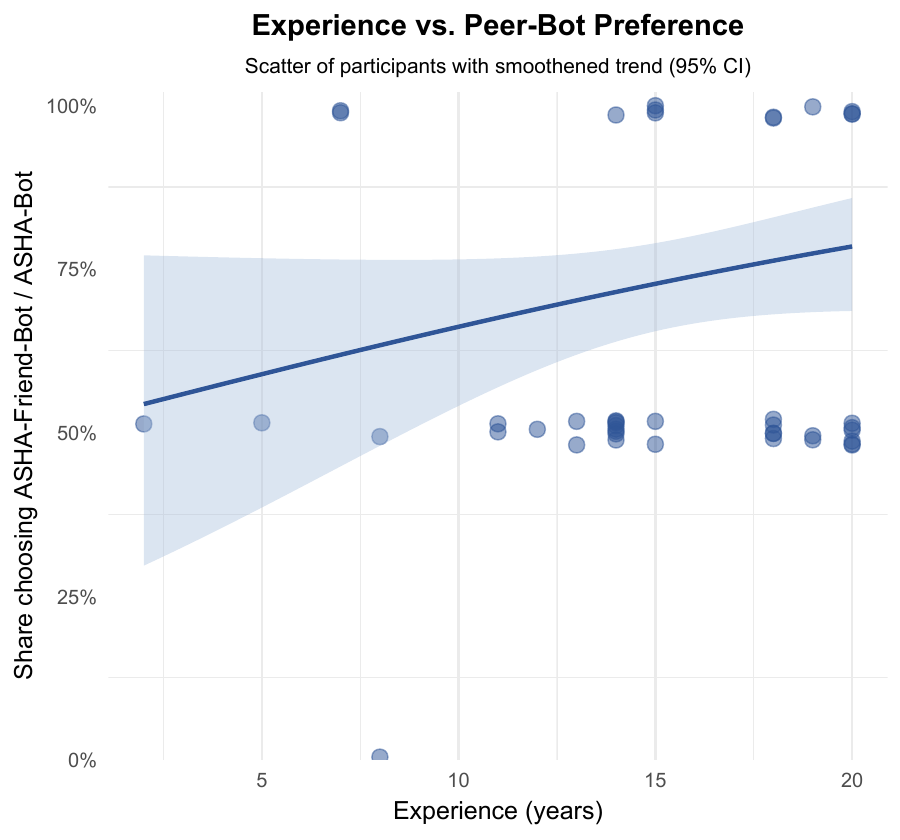}\label{fig:demog-exp}}
  \caption{Demographic gradients in peer-bot preference. Each point represents one ASHA's proportion of peer-oriented choices; smooths show descriptive trends.}
  \label{fig:demog-peer-pref}
\end{figure*}

\begin{figure*}[h]
  \centering
  \subfloat[Education vs.\ probability of following wrong advice (logistic smooth with 95\% CI)]{\includegraphics[width=0.45\linewidth]{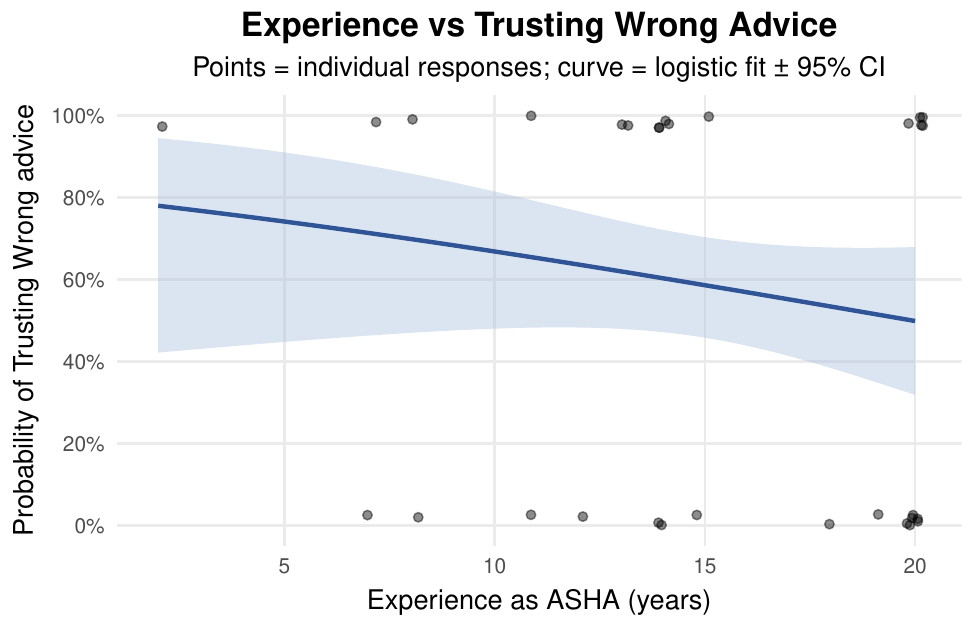}\label{fig:wrong-exp-overall}}
  \hfill
  \subfloat[Experience vs.\ probability of following wrong advice (logistic smooth with 95\% CI)]{\includegraphics[width=0.45\linewidth]{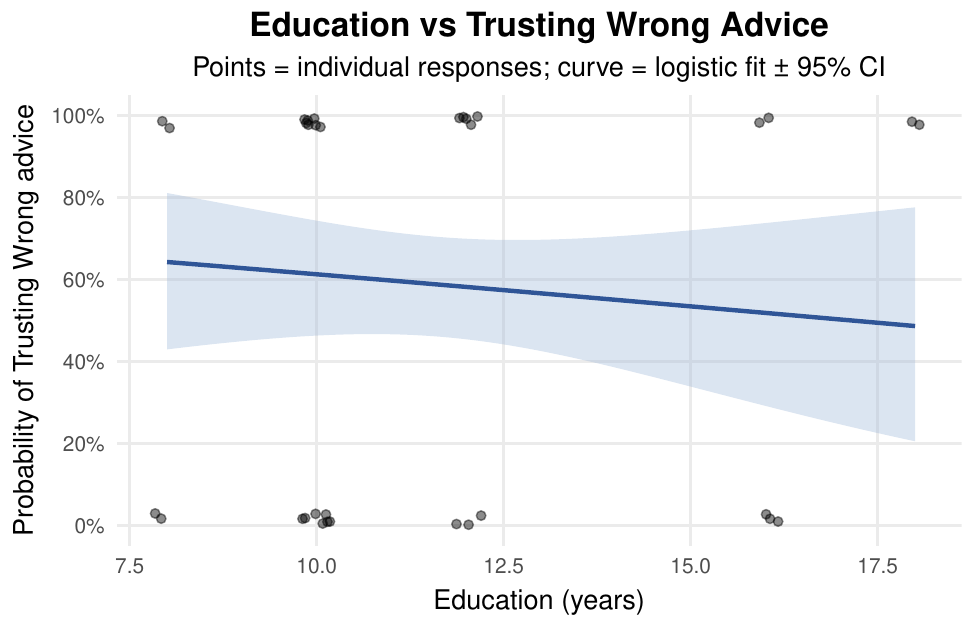}\label{fig:wrong-edu-overall}}
  \caption{Demographic trends in trusting the wrong advice given by a chatbot interface.}
  \label{fig:wrong-advice-demog}
\end{figure*}

\mobilehciadd{\subsection*{Appendix D: Qualitative Coding Process}}
\label{appendix:qualcoding}

\mobilehciadd{We analyzed the 61 interview transcripts using Braun and Clarke's reflexive thematic analysis~\cite{braun2006thematicanalysis}. We chose this approach because our goal was interpretive. We wanted to understand how ASHAs themselves reasoned about trust, framing, and feasibility, and to let themes emerge from their own words. The analysis had four phases. In the first phase, the first author transcribed all interviews from Hindi audio and translated them into English. Where a direct translation would lose meaning, we kept the original Hindi phrasing in brackets. For example, ``didi'' is a respectful term for an older sister or peer. In the second phase, the first author did line-by-line inductive coding on all 61 transcripts. This produced an initial codebook of 47 codes, grouped into five broad categories. In the third phase, all co-authors met across multiple sessions to combine related codes into higher-order themes. We drew on both the patterns in the data and on theories of normative framing (descriptive, narrative identity, and injunctive). In the fourth phase, we returned to the transcripts to verify each theme. We checked that each theme appeared across multiple participants. We discarded any theme that was driven by a single account. The themes reported in Section~\ref{sec:findings}, such as ``peer-like reassurance encouraged fast acceptance'' and ``social proof substituted for missing knowledge,'' came out of this process. Table~\ref{tab:codebook} lists the five categories, the codes within each, and an example quote from the interviews.}

\begin{table}[h]
\centering
\caption{Codebook of qualitative themes. Five categories with 47 codes in total emerged from inductive coding of 61 interview transcripts.}
\label{tab:codebook}
\small
\begin{tabular}{p{3.2cm} p{4.2cm} p{6.5cm}}
\toprule
\textbf{Category} & \textbf{Representative codes} & \textbf{Example quote and context} \\
\midrule
Affective responses to framing &
``feels like a peer or sister'';
``feels judgmental or scolding'';
``reads like a training session'';
``feels impersonal'';
``feels like a government circular'';
``approachable for sensitive topics''
&
``It felt like talking to another ASHA who understands.'' (P43, on ASHA-Friend-Bot) \\
\addlinespace
Decision rationales for trust and override &
``matched my training'';
``contradicted my training'';
``uncertain, deferred to source'';
``uncertain, deferred to peer endorsement'';
``checked against own experience'';
``fast acceptance without reflection'';
``paused to verify''
&
``I knew the right answer already, but when it came in that official style I slowed down and compared.'' (P27, on Health-Dept-Bot) \\
\addlinespace
Cross-framing comparisons &
``Friend-Bot vs. supervisor's scolding'';
``Health-Dept-Bot reads like a government circular'';
``ASHA-Bot reflects what is done locally'';
``Info-Bot feels generic''
&
``This was too formal and heavy. It felt like it was checking me instead of helping me.'' (P23) \\
\addlinespace
Professional practice and hierarchy &
``supervisor relationships'';
``PHC distance and access'';
``ANM consultations'';
``peer network as first line of consult'';
``training curriculum references'';
``accountability concerns''
&
``The supervisor is at the PHC, which is often a few kilometers away$\ldots$ I rarely get network.'' (P14) \\
\addlinespace
Contextual constraints &
``network unavailability in the field'';
``family resistance or aggression'';
``caste- or religion-based hesitancy'';
``travel distance'';
``time pressure during fieldwork'';
``literacy of recipient family''
&
``A Muslim woman was hesitant to get vaccinated because there were rumors $\ldots$ The answer was not in the training handbook.'' (P53) \\
\bottomrule
\end{tabular}
\end{table}

\end{document}